\documentclass[journal]{IEEEtran}
\usepackage[T1]{fontenc}
\ifCLASSINFOpdf
  \usepackage[pdftex]{graphicx}
  \usepackage[pdftex]{xcolor}
\else
  \usepackage[dvips]{graphicx}
\fi
\usepackage{array,amsmath,amsfonts,amssymb}
\usepackage{mathrsfs}
\usepackage{algorithm,algorithmic}
\usepackage{multirow, booktabs}
\usepackage{url}
\usepackage{cite}
\usepackage{accents}
\usepackage{booktabs}
\usepackage{multirow}
\usepackage[absolute,showboxes]{textpos}
\TPMargin{5pt}

\DeclareMathOperator*{\argmin}{argmin}

\newcommand{\romaone}{$\rm{\hspace{.18em}i\hspace{.18em}}$}
\newcommand{\romatwo}{$\rm{\hspace{.08em}ii\hspace{.08em}}$}

\newcommand{\copyrightstatement}{
\begin{textblock}{0.8}(0.1,0.01)
\noindent
\footnotesize
\copyright 2022 IEEE.
Personal use of this material is permitted.
Permission from IEEE must be obtained for all other uses, in any current or future media, including reprinting/republishing this material for advertising or promotional purposes, creating new collective works, for resale or redistribution to servers or lists, or reuse of any copyrighted component of this work in other works.
\end{textblock}}

\begin{document}
\copyrightstatement
\title{Online Phase Reconstruction via DNN-based \\ Phase Differences Estimation}

\author{Yoshiki Masuyama,~\IEEEmembership{Graduate Student Member,~IEEE,}
        Kohei Yatabe,~\IEEEmembership{Member,~IEEE,}
        Kento Nagatomo, \\
        and~Yasuhiro Oikawa,~\IEEEmembership{Member,~IEEE}
\thanks{Manuscript received May 8, 2022; revised Aug 31, 2022; revised Oct 12, 2022; accepted Oct 26, 2022.
\textit{(Corresponding author: Yoshiki Masuyama.)}}
\thanks{
Y. Masuyama is with the Department of Computer Science, Graduate School of Systems Design, Tokyo Metropolitan University, Hino, Tokyo 191-0065, Japan
(e-mail: masuyama-yoshiki@ed.tmu.ac.jp).

K. Yatabe is with the Department of Electrical Engineering and Computer Science, Tokyo University of Agriculture and Technology, Tokyo 184-8588, Japan
(e-mail: yatabe@go.tuat.ac.jp).

K. Nagatomo and Y. Oikawa are with the Department of Intermedia Art and Science, Waseda University, Tokyo 169-8555, Japan
(e-mail: jimijeffericking@akane.waseda.jp; yoikawa@waseda.jp).
}
}

\maketitle
\begin{abstract}
This paper presents a two-stage online phase reconstruction framework using causal deep neural networks (DNNs).
Phase reconstruction is a task of recovering phase of the short-time Fourier transform (STFT) coefficients only from the corresponding magnitude. 
However, phase is sensitive to waveform shifts and not easy to estimate from the magnitude even with a DNN.
To overcome this problem, we propose to use DNNs for estimating differences of phase between adjacent time-frequency bins.
We show that convolutional neural networks are suitable for phase difference estimation, according to the theoretical relation between partial derivatives of STFT phase and magnitude. 
The estimated phase differences are used for reconstructing phase by solving a weighted least squares problem in a frame-by-frame manner.
In contrast to existing DNN-based phase reconstruction methods, the proposed framework is causal and does not require any iterative procedure.
The experiments showed that the proposed method outperforms existing online methods and a DNN-based method for phase reconstruction.
\end{abstract}

\IEEEpeerreviewmaketitle

\section{Introduction}

\IEEEPARstart{P}{hase} reconstruction of short-time Fourier transform (STFT) coefficients is important for various audio technologies such as speech enhancement~\cite{Paliwal2011,LeRoux2013,Krawczyk2014,Gerkmann2015,Mowlaee2015,Mowlaee2016}, audio source separation~\cite{Magron2018,Takahashi2018,LeRoux2019,Wang2019,Masuyama2020b}, and text-to-speech synthesis~\cite{Takaki2017,Kaneko2017,Wang2017,Saito2018}.
As the structure of audio signals is apparent in the magnitude of STFT coefficients, ordinary methods for these technologies have focused on manipulating the magnitude.
After obtaining the magnitude, the corresponding phase is required to reconstruct a time-domain signal.
Although the phase of an observed signal is available in speech enhancement and audio source separation, it often causes artifacts and residual interference~\cite{Ephraim1984,Magron2015a}.
In text-to-speech synthesis, the magnitude is generated from linguistic features, and thus the phase is fully unavailable.
Hence, phase reconstruction of STFT coefficients is helpful for many applications.

\begin{figure}[t!]
  \centering
  \includegraphics[width=0.99\columnwidth]{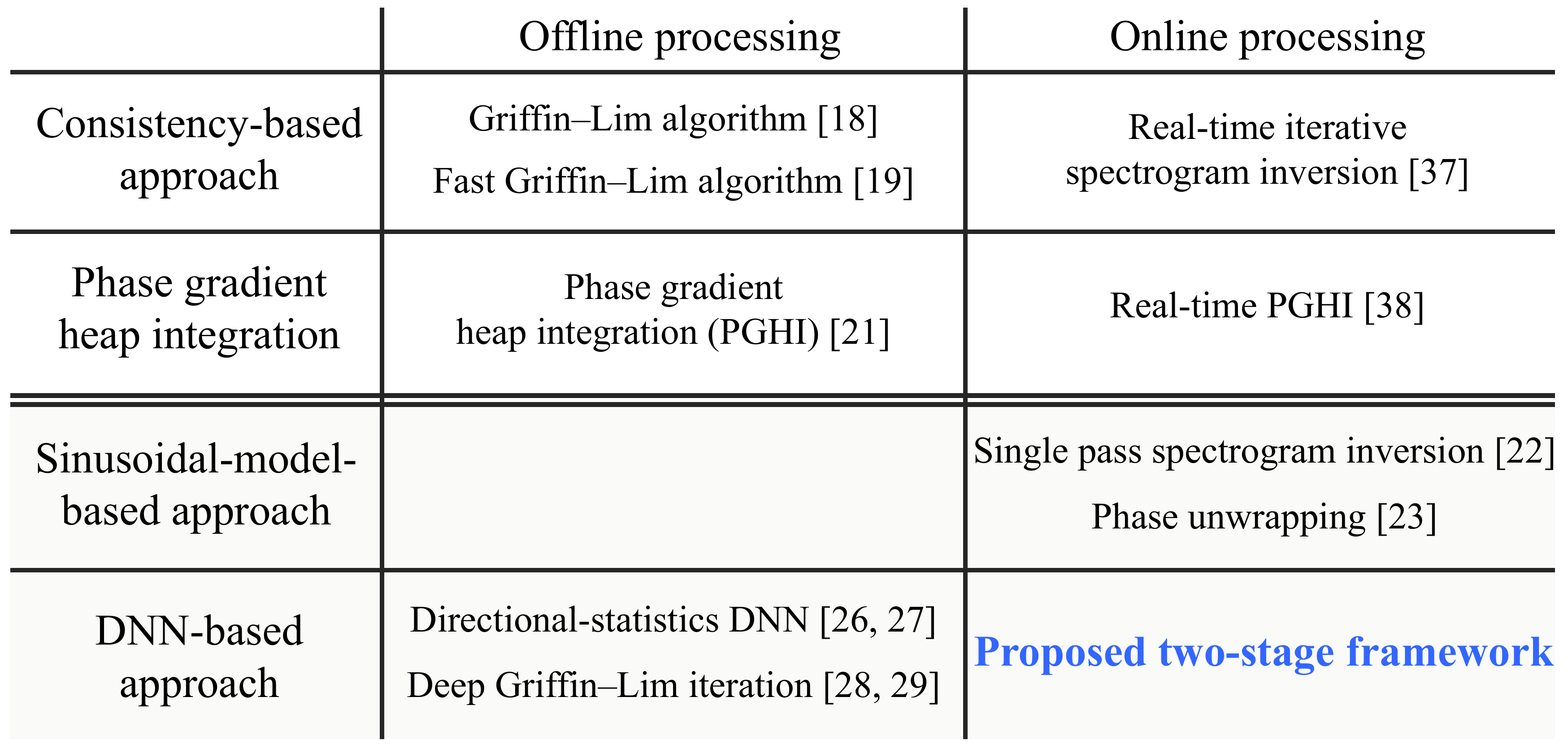}
  \caption{Comparison of offline and online phase reconstruction methods.
  Methods in the bottom half exploit the prior knowledge of the target signal.}
  \label{fig:overview}
\end{figure}

As summarized in Fig.~\ref{fig:overview}, various phase reconstruction methods have been studied,
such as consistency-based methods~\cite{Griffin1984,Perraudin2013,Masuyama2019a}, phase gradient heap integration (PGHI)~\cite{Prusa2017}, sinusoidal-model-based methods~\cite{Beauregard2015,Magron2015b}, and deep neural network (DNN)--based methods~\cite{Oyamada2018,Engel2019,Takamichi2018,Takamichi2020,Masuyama2019b,Masuyama2021,Masuyama2020a,Thieling2021,Thien2021}.
These methods can be divided into two categories: phase reconstruction with and without prior knowledge of a target signal.

As phase reconstruction methods without prior knowledge, consistency-based methods have been widely used~\cite{Griffin1984,Perraudin2013,Masuyama2019a}.
These methods are based on the relation among nearby STFT coefficients owing to the window and its overlapping nature.
To recover this relation, most consistency-based methods are given as iterative optimization algorithms.
Meanwhile, PGHI~\cite{Prusa2017} does not require such iterative procedures or the prior knowledge.
PGHI is based on a relation between the magnitude and phase of the STFT coefficients~\cite{Portnoff1979,Auger2012}; partial derivatives of phase can be analytically calculated from magnitude under some assumptions.
This relation enables PGHI to reconstruct phase by integrating phase derivatives.
As a result, PGHI has achieved promising results despite lacking any prior knowledge about the target signal.

In contrast, sinusoidal-model-based and DNN-based methods leverage prior knowledge about the target signal.
The assumption of sinusoidal-model-based methods is that the target signal consists of sinusoids~\cite{Beauregard2015,Magron2015b}.
This assumption allows one to approximate the phase derivative with respect to time by the frequency of each sinusoid.
Then, the phase can be reconstructed by integrating the approximated derivative over time.
However, such theoretically derived methods are only applicable to specific signals, and DNN-based methods are promising because DNNs can automatically learn prior knowledge from a training dataset.
Since some DNN-based methods have been successfully applied to phase reconstruction~\cite{Masuyama2021}, we also propose to use DNNs to exploit prior knowledge learned from the training dataset.

\begin{figure}[t!]
  \centering
  \includegraphics[width=0.99\columnwidth]{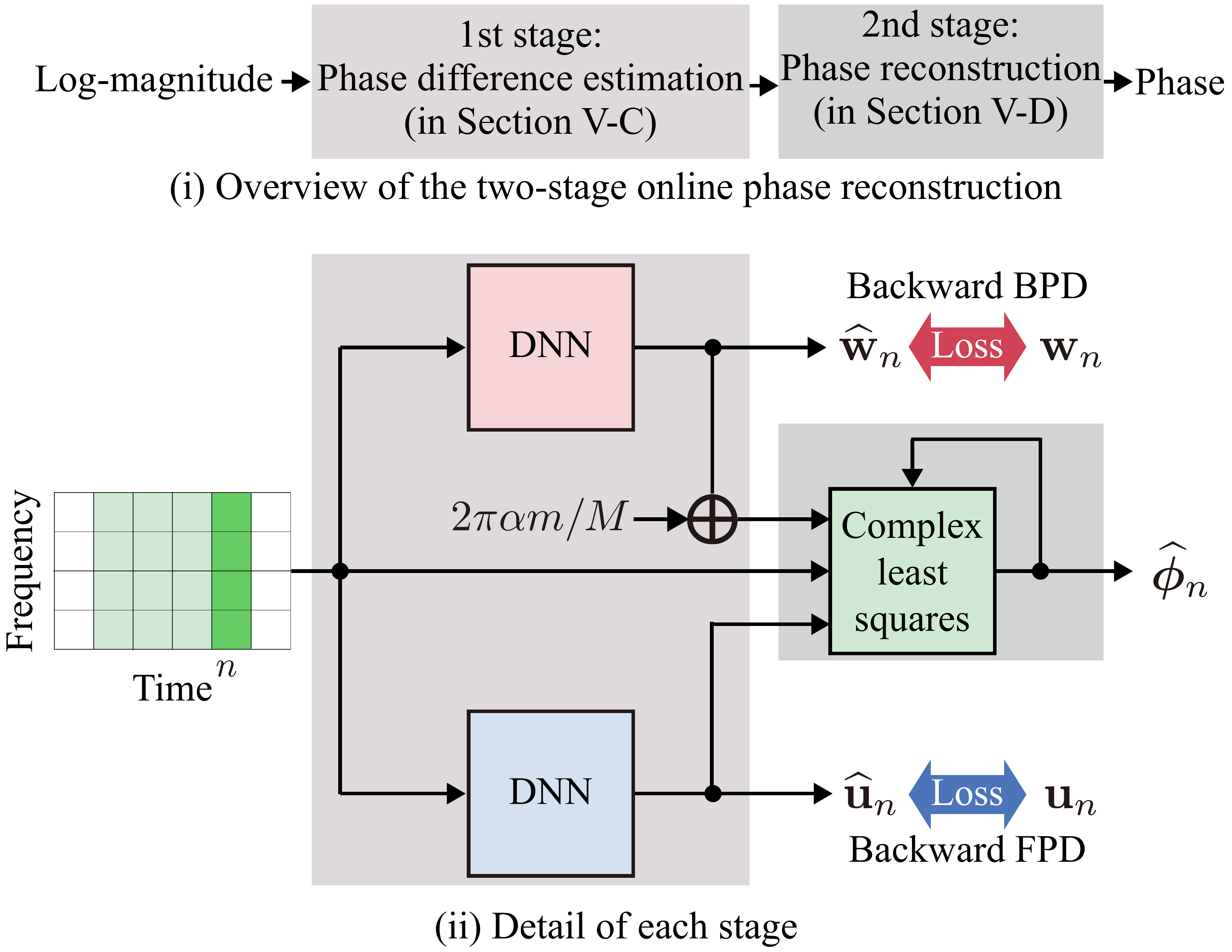}
  \caption{Illustration of the two-stage online phase reconstruction.
  As depicted in the top figure, the first stage estimates phase differences from the log-magnitude, and the second stage reconstructs phase from them.
  Instead of TPD, the DNN estimates a modified version of TPD called baseband phase delay (BPD) as shown in the bottom figure.
  The estimated BPD is converted to TPD by adding $2\pi \alpha m /M$.
  The second stage reconstructs the phase frame-by-frame by solving a weighted least squares problem of complex STFT coefficients.
  }
  \label{fig:proposed}
\end{figure}

While many phase reconstruction methods are offline algorithms, online phase reconstruction is highly desired in a wide range of applications, including incremental text-to-speech~\cite{Yanagita2019} and low-latency audio source separation~\cite{Magron2020}.
Therefore, except for the DNN-based methods, the aforementioned methods have been extended to the online setting~\cite{Zhu2007,Beauregard2015,Magron2015b,Prusa2016}.
A promising method is an extension of PGHI called real-time PGHI (RTPGHI)~\cite{Prusa2016}.
Although it outperforms the consistency-based method~\cite{Zhu2007} and the sinusoidal-model-based method~\cite{Beauregard2015}, there remains room for improvement.
First, the STFT phase-magnitude relation is valid only in the continuous setting, and thus estimated phase differences contain some errors in the discrete setting.
Second, to approximate the phase derivatives, the centered difference scheme used in PGHI is not allowable in the online setting without look-ahead frames.
Instead, RTPGHI uses the backward difference that results in lower performance than the original PGHI.
The strong modeling capability of DNNs can improve the estimation accuracy of phase differences.

In this paper, we propose a DNN-based online phase reconstruction framework.
As illustrated in Fig.~\ref{fig:proposed}, the proposed framework consists of two stages: (\romaone) estimating the phase differences from the magnitude and (\romatwo) reconstructing the phase from the estimated phase differences.
First, we estimate the phase differences with respect to time (TPD) and frequency (FPD) by using causal DNNs.
This DNN-based estimation is expected to be robust to the mismatch of the STFT phase-magnitude relation by leveraging the prior knowledge acquired from a training dataset.
Second, we recurrently reconstruct phase from the estimated differences.
To handle the phase differences efficiently, we treat them as the ratios of complex STFT coefficients.
Then, the phase is reconstructed by solving a weighted least-squares problem of STFT coefficients.
Through several experiments, we confirmed the effectiveness of the proposed two-stage framework compared to existing online phase reconstruction methods.

Note that this paper is related to our conference paper~\cite{Masuyama2020a}, in which we developed the basic concept of the two-stage phase reconstruction framework in the offline setting.
In this paper, we extend it to an online method with improvement on all components, i.e., both the first and the second stages.
The contributions of this paper are summarized as follows:
\begin{itemize}
    \item proposing an online phase reconstruction framework using causal DNNs, while our previous work \cite{Masuyama2020a} focused on the offline setting;
    \item applying convolutional neural networks (CNNs) to estimate phase differences, which is motivated by the STFT phase-magnitude relation;
    \item presenting a novel method for reconstructing phase from its differences by solving the weighted least squares problem of complex STFT coefficients;
    \item investigating and comparing the performance of various online phase reconstruction methods.
\end{itemize}

The rest of the paper is organized as follows.
In Section~\ref{sec:pr-formulation}, offline and online phase reconstruction problems are formulated.
Section~\ref{sec:PGHI} explains the STFT phase-magnitude relation, PGHI, and its online extension, RTPGHI.
DNN-based phase reconstruction methods are also reviewed.
The proposed two-stage framework for DNN-based online phase reconstruction is introduced in Section~\ref{sec:proposed}.
In Section~\ref{sec:exp}, the proposed method is compared with various online phase reconstruction methods, and then the effectiveness of both stages is investigated.
Finally, Section~\ref{sec:conclusion} concludes this paper.

\section{Problem formulation}
\label{sec:pr-formulation}

STFT%
\footnote{
In the literature of audio signal processing, the transform defined by \eqref{eq:stft-discrete} is commonly called STFT, while it is called the discrete Gabor transform in other communities~\cite{Feichtinger1998,Groechenig2001}.
In this paper, we use the term STFT according to the former literature.%
}
of a discrete signal $\boldsymbol{\chi}$ with respect to a real symmetric window $\mathbf{g}$ of length $L$ is defined as
\begin{align}
    X[m,n] = \sum_{l=-\lfloor L/2 \rfloor}^{\lfloor L/2 \rfloor} \chi[l + \alpha n] \, g[l] \, \mathrm{e}^{- 2 \pi \mathrm{i} l m /M},
    \label{eq:stft-discrete}
\end{align}
where $X[m,n]$ is the $(m,n)$th entry of the STFT coefficients,
$\mathrm{i}$ is the imaginary unit, $\alpha$ is a time shifting step,
$n = 0, \ldots, N-1$ and $m = 0, \ldots, M-1$ are the time-frame and frequency indices, respectively.
These symbols used in this section are summarized in Table~\ref{tab:notations-pr-formulation}.
Let us denote the magnitude and phase of the STFT coefficients $\mathbf{X}$ by $\mathbf{A}$ and $\boldsymbol{\Phi}$, respectively:
\begin{align}
    A[m,n] &= |X[m,n]| \\
    \Phi[m,n] &= \mathrm{Arg}(X[m,n]),
\end{align}
where $\mathrm{Arg}(\cdot)$ returns the principal value of the complex-argument of its input.

We consider the task of phase reconstruction that aims at estimating the target phase $\boldsymbol{\Phi}$ while allowing the ambiguity of $2\pi$.
That is, our objective is to construct the a mapping $\mathscr{F}(\cdot)$ such that:
\begin{align}
    \widehat{\boldsymbol{\Phi}} &= \mathscr{F}(\mathbf{A}), \label{eq:pr} \\
    \boldsymbol{\Phi} &\approx \widehat{\boldsymbol{\Phi}} +2\pi\mathbf{N},
\end{align}
where $\mathbf{N}\in\mathbb{Z}^{M \times N}$ is an arbitrary integer-valued array.

Offline phase reconstruction uses the STFT magnitude at all time-frames as in \eqref{eq:pr}, which can be written as follows:
\begin{equation}
    \widehat{\boldsymbol{\Phi}} = \mathscr{F}(\mathbf{a}_0, \ldots, \mathbf{a}_{N-1}), \label{eq:offline-pr}
\end{equation}
where $\mathbf{a}_n = [A[0,n], \ldots, A[M-1,n]]^\mathsf{T}$, and $(\cdot)^\mathsf{T}$ denotes the transpose.
However, \eqref{eq:offline-pr} is not applicable to the online setting.
In real-time applications, we should estimate the phase at each time-frame only from the magnitudes up to the current time-frame and few look-ahead frames:
\begin{equation}
    \widehat{\boldsymbol{\phi}}_n = \mathscr{F}(\ldots, \mathbf{a}_{n}, \ldots, \mathbf{a}_{n+N_\mathrm{LA}}),
    \label{eq:online-pr}
\end{equation}
where $\widehat{\boldsymbol{\phi}}_n = [\widehat{\Phi}[0,n], \ldots, \widehat{\Phi}[M-1,n]]^\mathsf{T}$, and $N_\mathrm{LA} \in \mathbb{N}$ is the number of look-ahead frames.
The number of time-frames that affect the current output depends on the map $\mathscr{F}(\cdot)$.
When using a causal CNN~\cite{Oord2016}, it depends on the receptive field of the CNN.
Meanwhile, when using a recurrent neural network (RNN), the output at the current time-frame implicitly depends on the magnitudes at all the past time-frames.
In this paper, a system is said to be causal if it does not require future information to compute its current output, i.e., $N_\mathrm{LA}=0$.

\begin{table}[t]
\caption{List of Symbols Used in Section~\ref{sec:pr-formulation}}
\label{tab:notations-pr-formulation}
\vspace{-4pt}
\begin{center}
\scalebox{0.9}{
\begin{tabular}{c|c}
\hline
\hline
\multicolumn{2}{c}{Variables} \\
\hline
${\chi}[l]$ & The $l$th sample of a discrete time-domain signal \\
${g}[l]$ & The $l$th sample of a window used in STFT \\
$X{[m, n]}$ & Complex STFT coefficient at the $(m,n)$th bin \\
$A{[m, n]}$ & STFT magnitude given by $|X{[m, n]}|$ \\
$\Phi{[m, n]}$ & STFT phase given by $\mathrm{Arg}(X{[m, n]})$ \\
\hline
\multicolumn{2}{c}{Notations with accents and subscripts} \\
\hline
$\widehat{(\cdot)}$ & Estimate of its input \\
\multirow{2}{*}{$(\cdot)_n$} & Vector of its input at the $n$th time-frame, \\
& e.g., $\mathbf{x}_n = [X[0,n], \ldots, X[M-1,n]]^\mathsf{T}$ \\
\hline
\multicolumn{2}{c}{Maps} \\
\hline
$\mathscr{F}(\cdot)$ & Mapping from the magnitude to the phase \\
$\mathrm{Arg}(\cdot)$ & Mapping from a complex scalar to its principal argument \\
\hline
\hline
\end{tabular}
}
\end{center}
\end{table}

\section{Related Works}
\label{sec:PGHI}

In this section, after explaining PGHI and RTPGHI, we review the DNN-based phase reconstruction methods.
The symbols used in this section are listed in Table~\ref{tab:notations-pghi}.

\begin{table}[t]
\caption{List of Symbols Used in Section~\ref{sec:PGHI}}
\label{tab:notations-pghi}
\vspace{-4pt}
\begin{center}
\scalebox{0.9}{
\begin{tabular}{c|c}
\hline
\hline
\multicolumn{2}{c}{Variables} \\
\hline
$y$ & $L^2$ function as a signal \\
$h$ & $L^2$ window function \\
$Y$ & Continuous STFT of the function $y$ \\
$\mathcal{A}$ & Magnitude of $Y$ \\
$\varphi$ & Phase of $Y$ \\
$\widetilde{A}[m,n]$ & Log-magnitude of the discrete STFT coefficient \\
${V}_{\mathrm{c}} [m,n]$ & Phase derivative for time defined for the $(m,n)$th point \\
${U}_{\mathrm{c}} [m,n]$ & Phase derivative for frequency defined for the $(m,n)$th point \\
${V} [m,n]$ & Backward phase difference for time (TPD) \\
${U} [m,n]$ & Backward phase difference for frequency (FPD) \\
\hline
\multicolumn{2}{c}{Maps} \\
\hline
$\mathcal{F}_{\boldsymbol{\theta}}(\cdot)$ & DNN for estimating phase from the given magnitude \\
$\mathcal{L}(\cdot, \cdot)$ & Periodic loss function \\
\hline
\hline
\end{tabular}
}
\end{center}
\end{table}

\subsection{Phase Gradient Heap Integration (PGHI)}

PGHI is a non-iterative phase reconstruction method based on the STFT phase-magnitude relation derived from the definition of continuous STFT~\cite{Prusa2017}.
In the continuous setting, STFT of a function $y \in L^2(\mathbb{R})$ with respect to a window function $h \in L^2(\mathbb{R})$ is defined as 
\begin{align}
    Y(f, t) &= \int_\mathbb{R} y (\tau + t) \, h(\tau) \, \mathrm{e}^{- 2 \pi \mathrm{i} f \tau} \mathrm{d} \tau \nonumber \\
    &= \mathcal{A} (f,t) \, \mathrm{e}^{i \varphi (f,t)},
\end{align}
where $\mathcal{A}$ and $\varphi$ represent the magnitude and phase, respectively.
Let us define the Gaussian window as follows:
\begin{equation}
    h(t) = \left( \frac{2}{\sigma^2} \right)^{1/4} \mathrm{e}^{-\pi t^2 / \sigma^2},
\end{equation}
where $\sigma$ is a parameter of the Gaussian window.
When using the Gaussian window for STFT, both magnitude and phase are partially differentiable with respect to both time and frequency.
In addition, the following phase-magnitude relation of STFT can be derived~\cite{Prusa2017}:
\begin{align}
    \frac{\partial}{\partial t} \varphi (f,t)
    &= \frac{1}{\sigma^2}\frac{\partial}{\partial f} \log(\mathcal{A} (f,t)) + 2\pi f,
    \label{eq:partialtime_c}\\
    \frac{\partial}{\partial f} \varphi (f,t)
    &= - \sigma^2 \frac{\partial}{\partial t} \log(\mathcal{A} (f,t)). \label{eq:partialfreq_c}
\end{align}
This relation indicates that the phase derivatives can be analytically calculated from the corresponding log-magnitude.
We can thus reconstruct the phase by integrating its gradient up to the global constant phase.

PGHI exploits the relations in \eqref{eq:partialtime_c} and \eqref{eq:partialfreq_c} to compute the phase gradient in the discrete setting, where STFT is defined as \eqref{eq:stft-discrete}.
In PGHI, phase gradient is approximated by using the second order centered differences of log-magnitude:
\begin{align}
    \!\!\widehat{V}_{\mathrm{c}} [m,n] &= \frac{\alpha M}{2\beta} (\widetilde{A}[m+1,n] - \widetilde{A}[m-1,n]) + \frac{2\pi \alpha m}{M}, \!\!
    \label{eq:tdiff-ana}\\
    \!\!\widehat{U}_{\mathrm{c}} [m,n] &= - \frac{\beta}{2\alpha M} (\widetilde{A}[m,n+1] - \widetilde{A}[m,n-1]),
    \label{eq:fdiff-ana}
\end{align}
where $\widetilde{A}[m,n] = \log({A}[m,n])$, and $\beta$ is a constant depending on the window.
Note that $\widehat{V}_{\mathrm{c}}[m,n]$ and $\widehat{U}_{\mathrm{c}}[m,n]$ approximate phase derivatives sampled on the time-frequency (T-F) grid.

\begin{figure}[t!]
  \centering
  \includegraphics[width=0.9\columnwidth]{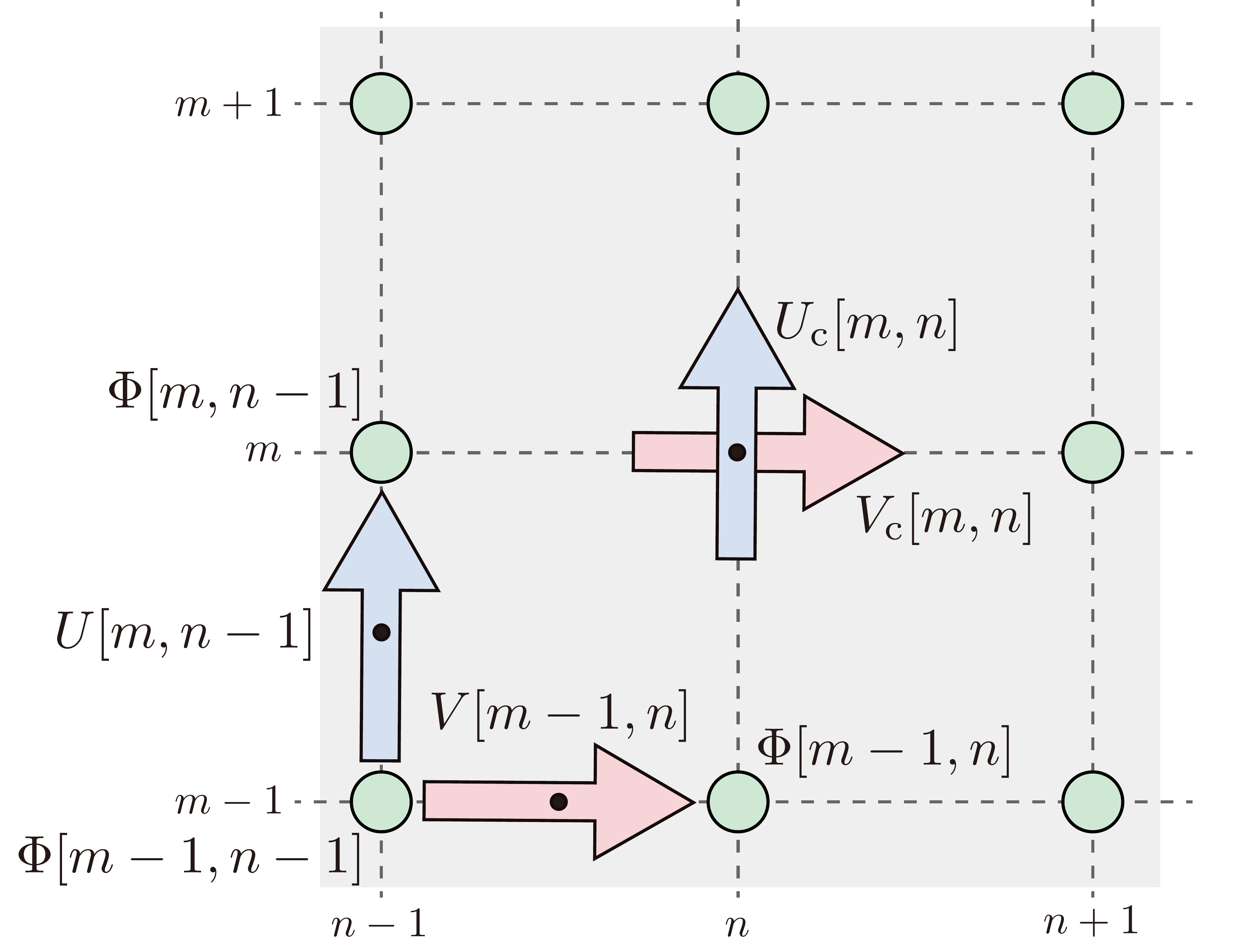}
  \caption{Illustration of the phase derivatives and the backward phase differences.
  Green circles correspond to phases at the T-F grids. Red and blue arrows indicate phase differences with respect to time and frequency, respectively.}
  \label{fig:v-and-u}
\end{figure}

The phase is reconstructed by numerically integrating the backward phase differences.
Since there are multiple possible paths for the integration, PGHI adaptively chooses one of the following four integration paths:
\begin{align}
    \widehat{\Phi}[m,n] &= \widehat{\Phi}[m,n-1] + \widehat{V} [m,n],
    \label{eq:pr-rule-time-forward} \\
    \widehat{\Phi}[m,n] &= \widehat{\Phi}[m,n+1] - \widehat{V} [m,n+1],
    \label{eq:pr-rule-time-backward}\\
    \widehat{\Phi}[m,n] &= \widehat{\Phi}[m-1,n] + \widehat{U}[m,n],
    \label{eq:pr-rule-freq-forward}\\
    \widehat{\Phi}[m,n] &= \widehat{\Phi}[m+1,n] - \widehat{U}[m+1,n],
    \label{eq:pr-rule-freq-backward}
\end{align}
where $\widehat{V}[m,n]$ and $\widehat{U}[m,n]$ are approximate backward TPD and FPD, respectively.
They are given by averaging the estimated phase gradient in \eqref{eq:tdiff-ana} and \eqref{eq:fdiff-ana}:
\begin{align}
    \widehat{V}[m,n] &= \frac{\widehat{
V}_{\mathrm{c}} [m,n] + \widehat{
V}_{\mathrm{c}} [m,n-1]}{2},
    \label{eq:tdiff-ave-ana} \\
    \widehat{U}[m,n] &= \frac{\widehat{
U}_{\mathrm{c}} [m,n] + \widehat{
U}_{\mathrm{c}} [m-1,n]}{2}.
    \label{eq:fdiff-ave-ana}
\end{align}
Note that the oracle backward TPD and FPD are given by
\begin{align}
    {V}[m,n] &=\Phi[m,n] - \Phi[m,n-1],
    \label{eq:tdiff-oracle-ana} \\
    U[m,n] &= \Phi[m,n] - \Phi[m-1,n],
    \label{eq:fdiff-oracle-ana}
\end{align}
respectively.
The relation between the phase derivatives and the backward phase differences are illustrated in Fig.~\ref{fig:v-and-u}.
While the former is defined for every T-F points, the latter is defined as the relation between adjacent T-F bins.

In the numerical integration, PGHI omits phase at a T-F bin whose magnitude is small because phase differences should be unreliable at such T-F bins.
Instead, random phase is assigned to such T-F bins for simplicity.
We refer the reader to the paper~\cite{Prusa2017} and codes%
\footnote{
PGHI and RTPGHI are implemented in the phase retrieval toolbox (\texttt{PHASERET}):
\url{http://ltfat.github.io/phaseret/}~\cite{Prusa2017a}.
}
for more details of implementation.

\subsection{Real-time PGHI (RTPGHI)}
\label{sec:RTPGHI}

RTPGHI is an online extension of PGHI~\cite{Prusa2016}.
When allowing one look-ahead frame, i.e., $N_\mathrm{LA}=1$ in \eqref{eq:online-pr}, the phase gradient can be approximated by the centered difference scheme as in \eqref{eq:tdiff-ana} and \eqref{eq:fdiff-ana}.
However, the centered difference in \eqref{eq:fdiff-ana} is not applicable to the causal setting (i.e., $N_\mathrm{LA}=0$) because $\widetilde{A}[m,n+1]$ is not accessible.
Hence, RTPGHI approximates the phase derivative with respect to the frequency by using the second order backward time-difference of the log-magnitude:
\begin{align}
    \widehat{U}_{\mathrm{c}} [m,n] &= - \frac{\beta}{2\alpha M} (3\widetilde{A}[m,n]) - \nonumber \\ & \hspace{23pt} 4\widetilde{A}[m,n-1] + \widetilde{A}[m,n-2]).
    \label{eq:causal}
\end{align}
Then, the phase is reconstructed via one of the integration paths except \eqref{eq:pr-rule-time-backward} because $\widehat{\Phi}[m,n+1]$ is not available.

Although RTPGHI achieved promising results, it has some limitations.
First, the phase-magnitude relation in \eqref{eq:partialtime_c} and \eqref{eq:partialfreq_c} is only valid for the continuous case, and hence the estimated TPD and FPD in \eqref{eq:tdiff-ana} and \eqref{eq:fdiff-ana} contain errors in the discrete setting.
Moreover, this relation assumes the use of the Gaussian window which has infinite support in the time domain.
Such a window is not allowed in real-time applications.
Second, the experimental results in \cite{Prusa2016} showed that the second order backward difference approximation in \eqref{eq:causal} degrades the quality of the reconstructed signals from that of the centered difference in \eqref{eq:fdiff-ana}.

\subsection{DNN-based Phase Reconstruction}
\label{sec:DNN-based PR}

DNN-based phase reconstruction has gained increasing attention because of strong modeling capability of DNNs~\cite{Oyamada2018,Engel2019,Takamichi2018,Takamichi2020,Masuyama2019b,Masuyama2020a, Masuyama2021,Thieling2021,Thien2021}.
A DNN-based method can handle various signals by learning prior knowledge from a training dataset.
A straightforward approach is to model the map $\mathscr{F}(\cdot)$ in \eqref{eq:pr} by a DNN. 
When training such a DNN, we should consider the periodic nature of the target phase $\Phi[m,n]$ because phase is given as a complex-argument.
Ordinary loss functions for a regression problem, including the mean squared error, are not suitable for training in such a situation.

To address this issue, several approaches have been presented.
One approach uses a DNN to estimate complex STFT coefficients $\mathbf{X}$ instead of their phase $\boldsymbol{\Phi}$~\cite{Oyamada2018,Masuyama2019b,Masuyama2021}.
Another approach quantizes the target phases and estimates their indices~\cite{Takahashi2018,LeRoux2019}.
As a result of recasting the regression problem as a classification problem, the periodic nature of the phase is circumvented.
Neither approach directly estimates the phase to avoid dealing with a circular variable.

A periodic loss function has been proposed to train a DNN that directly estimates the continuous circular phase~\cite{Takamichi2018,Takamichi2020}.
In this approach, a DNN $\mathcal{F}_{\boldsymbol{\theta}}(\cdot)$ directly estimates the phase:
\begin{equation}
\widehat{\boldsymbol{\Phi}} = \mathcal{F}_{\boldsymbol{\theta}}(\mathbf{A}),
\end{equation}
where $\boldsymbol{\theta}$ is a set of parameters of the DNN.
To measure the error between the target phase $\Phi[m,n]$ and the estimated phase $\widehat{\Phi}[m,n]$, a periodic loss function satisfying
\begin{align}
    \mathcal{L}(\phi, \widehat{\phi}) = \mathcal{L}(\phi, \widehat{\phi} + 2\pi b)
\end{align}
is considered, where $b$ is an arbitrary integer.
For instance, the negative cosine loss function is given by
\begin{equation}
    \mathcal{L}_\text{cos}(\phi, \widehat{\phi}) = - \cos (\phi - \widehat{\phi}). \label{eq:vm-loss}
\end{equation}%
By using the periodic loss function, a DNN for estimating the continuous phase is trained as follows:
\begin{align}
    \min_{\boldsymbol{\theta}} \,\, \sum_{m=0}^{M-1} \sum_{n=0}^{N-1} \mathcal{L}_\mathrm{cos}(\Phi[m,n], \mathcal{F}_{\boldsymbol{\theta}}(\mathbf{A})[m,n]),
    \label{eq:takamichi}
\end{align}
where we omit the summation over the training dataset because the DNN treats each pair of $\mathbf{A}$ and $\boldsymbol{\Phi}$ separately.
The estimated phase has the ambiguity of $2\pi$ due to the use of the periodic loss function.
This ambiguity is not a problem when calculating the complex STFT coefficients with the given magnitude as $A[m,n] \exp(\mathrm{i}\Phi[m,n])$.
The direct phase estimation with a DNN is still hard because a small perturbation of magnitude might imply a large phase difference.

\section{Proposed Online Phase Reconstruction}
\label{sec:proposed}

In this section, we propose a DNN-based online phase reconstruction framework that consists of two stages.
Section \ref{sec:motivation} shows the motivation of the two-stage framework.
Then, its overview is introduced in Section \ref{sec:concept}.
The detail of each stage is explained in Sections~\ref{sec:stage1} and \ref{sec:stage2}, respectively.
The weighting rule in the second stage is presented in Section~\ref{sec:weight-rule}.
The symbols used in this section are summarized in Table~\ref{tab:Notations}.

\begin{table}[t]
\caption{List of Symbols Used in Section~\ref{sec:proposed}}
\label{tab:Notations}
\vspace{-4pt}
\begin{center}
\scalebox{0.9}{
\begin{tabular}{c|c}
\hline
\hline
\multicolumn{2}{c}{Variables} \\
\hline
$W{[m, n]}$ & Backward baseband phase difference (BPD) \\
$\mathfrak{V}[m,n]$ & Ratio between STFT coefficients at the adjacent time-frames \\
$\mathfrak{U}[m,n]$ & Ratio between STFT coefficients at the adjacent frequency-bins \\
$\boldsymbol{\Psi}_n$ & Feature matrix for estimating the phase differences by DNNs \\
$\boldsymbol{\Lambda}_n$ & Diagonal matrix  representing the reliability of estimated TPD \\
$\boldsymbol{\Gamma}_n$ & Diagonal matrix  representing the reliability of estimated FPD \\
\hline
\multicolumn{2}{c}{Maps} \\
\hline
$\mathcal{W}(\cdot)$ & Wrapping operator \\
\!\!\!\!$\mathcal{G}_{\boldsymbol{\theta}_\mathrm{time}}(\cdot)$\!\!\!\! & DNN for estimating BPD from the given magnitude \\
\!\!\!\!$\mathcal{H}_{\boldsymbol{\theta}_\mathrm{freq}}(\cdot)$\!\!\!\! & DNN for estimating FPD from the given magnitude\\
$\mathcal{P}(\cdot)$ & Autoregressive map for reconstructing phase from TPD and FPD \\
$\mathtt{Arg}(\cdot)$ & \!\!Element-wise map from complex scalars to their principal arguments\!\! \\
\hline
\hline
\end{tabular}
}
\end{center}
\end{table}

\subsection{Motivation: Sensitivity of Phase to Waveform Shift}
\label{sec:motivation}

A DNN-based phase reconstruction method in \cite{Takamichi2018} is formulated as $\widehat{\boldsymbol{\Phi}} = \mathcal{F}_{\boldsymbol{\theta}}(\mathbf{A})$.
When training such a DNN in a supervised manner, not only the periodic nature of the phase but also the sensitivity to waveform shifts becomes a problem.
Considering the Fourier transform, its phase is sensitive to waveform shifts, while its magnitude is shift-invariant.
This is approximately true for STFT when waveform shifts are small.
Furthermore, when the sign of a time domain signal is inverted, the corresponding STFT phase is shifted by $\pi$ without changing magnitude.
It is thus difficult to completely determine the phase from given magnitude%
\footnote{
STFT magnitude is not completely shift-invariant, and thus the phase can be reconstructed except for the ambiguity of $\pm \pi$ under several assumptions~\cite{Candes2013,Waldspurger2015}.
However, these phase reconstruction methods require excessive computation and are not suitable for the online setting.
}.

\begin{figure}[t!]
  \centering
  \includegraphics[width=0.99\columnwidth]{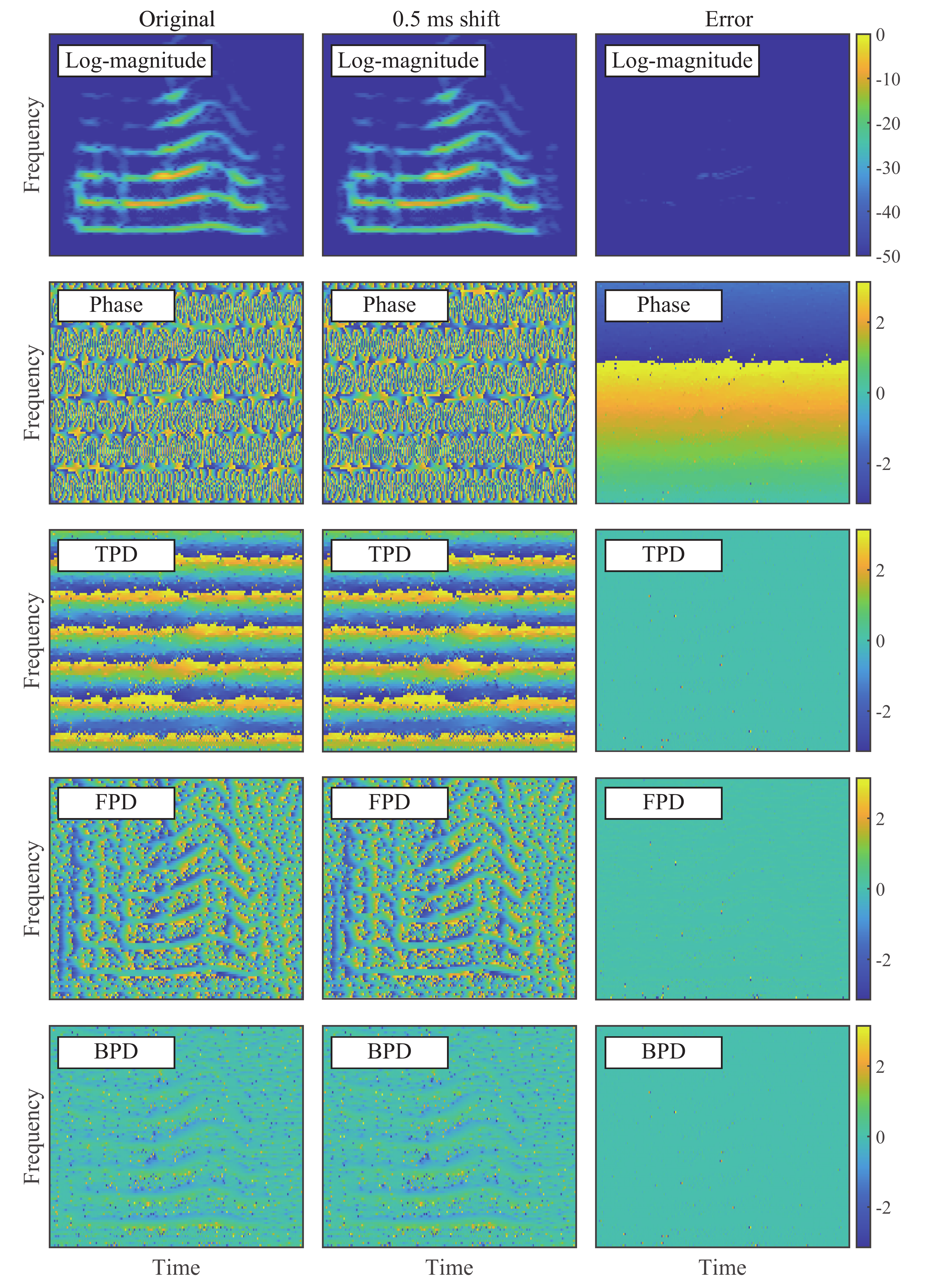}
  \caption{Examples of the STFT magnitude, phase, and backward phase differences of an utterance and that shifted by $0.5$ ms.
  The rightmost column shows the errors between the original and shifted ones.}
  \label{fig:shift-effect}
\end{figure}

The effects of a waveform shift on STFT magnitude and phase are depicted in Fig.~\ref{fig:shift-effect}.
We used an utterance in the LJ speech dataset%
\footnote{The LJ speech dataset is available in online: \url{https://keithito.com/LJ-Speech-Dataset/}.}
and that shifted by $0.5$ ms.
Their TPD and FPD are depicted in the third and fourth rows where we wrap them by using the following wrapping operator:
\begin{equation}
    \mathcal{W}(\Phi) = \mathrm{Arg}(\mathrm{e}^{\mathrm{i} \Phi}).
\end{equation}
Phase of the shifted signal is noticeably different from the original one even though the magnitude remained almost the same.
We thus expect that the phase itself is not easy to estimate from the magnitude.
In contrast, according to the third and fourth rows of Fig.~\ref{fig:shift-effect}, phase differences, TPD and FPD, were robust to the waveform shift.

The harmonic structure is apparent in the FPD but vague in the TPD. We thus modify TPD to BPD~\cite{Krawczyk2014}%
\footnote{
BPD (baseband phase delay) was introduced in a sinusoidal-model-based phase reconstruction~\cite{Krawczyk2014} and has also been used in DNN-based phase reconstruction~\cite{Thieling2021}.
This DNN-based phase reconstruction method uses the BPD to normalize the distribution of TPD.
In Section~\ref{sec:stage1}, we will show the importance of the modification in \eqref{eq:bpd} especially with CNNs.
}:
\begin{equation}
W[m,n] = \mathcal{W}\left(V[m,n] - \frac{2 \pi \alpha m}{M}\right). \label{eq:bpd}    
\end{equation}
As depicted in the bottom row of Fig.~\ref{fig:shift-effect}, the harmonic structure is more clear in BPD than in TPD.
We thus expect that BPD and FPD are easier to estimate by DNNs, where this expectation will be experimentally confirmed in Section~\ref{sec:exp-vs-direct}.

\subsection{Overview: Two-stage Online Phase Reconstruction}
\label{sec:concept}

\begin{figure}[t!]
  \centering
  \includegraphics[width=0.99\columnwidth]{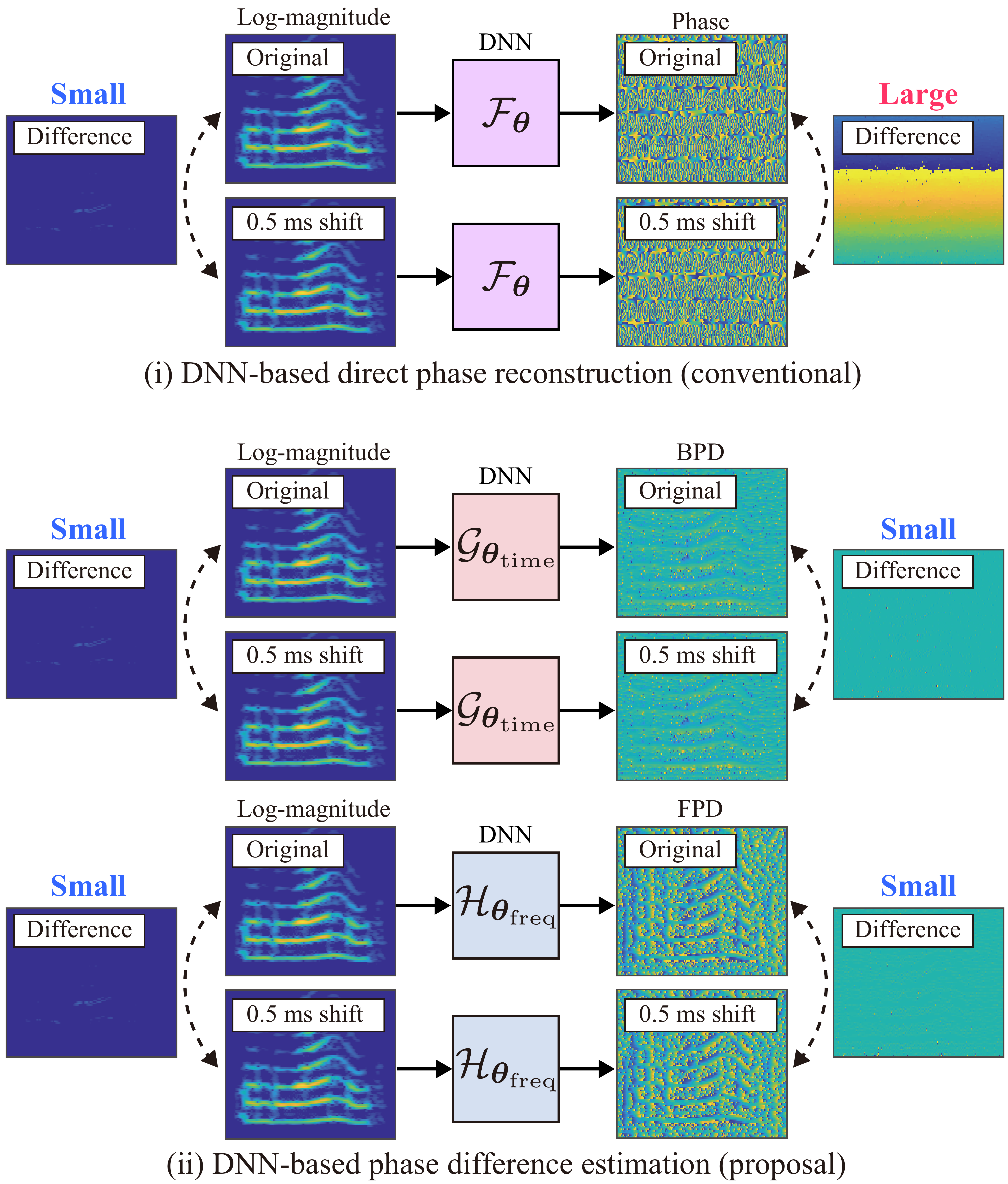}
  \caption{Comparison between (\romaone) the existing DNN-based direct phase reconstruction and (\romatwo) the first stage of the proposed framework.
  Although the estimation is performed separately for each time-frame, magnitude and phase differences at all time-frames are shown for visibility.
  }
  \label{fig:proposed-framework}
\end{figure}

The example in the previous subsection suggested that directly estimating phase is difficult because a DNN must connect small changes in magnitude to large differences in phase, as illustrated in Fig.~\ref{fig:proposed-framework}-(\romaone).
Such an unstable map is not easy to model by a DNN.
In contrast, we use DNNs to estimate BPD and FPD as depicted in Fig.~\ref{fig:proposed-framework}-(\romatwo).
Since a small change in log-magnitude results in small changes in BPD and FPD, the maps from the log-magnitude to the phase differences should be easily modeled by DNNs.
After estimating BPD and FPD, the phase is reconstructed in a frame-by-frame manner.
This two-stage phase reconstruction is summarized in Fig.~\ref{fig:proposed}.

At the first stage, causal DNNs $\mathcal{G}_{\boldsymbol{\theta}_\mathrm{time}}(\cdot)$ and $\mathcal{H}_{\boldsymbol{\theta}_\mathrm{freq}}(\cdot)$ estimate BPD $\mathbf{w}_n \in \mathbb{R}^{M}$ and FPD $\mathbf{u}_n \in \mathbb{R}^{M-1}$ at the $n$th frame, respectively, as follows:
\begin{align}
    \widehat{\mathbf{w}}_n &= \mathcal{G}_{\boldsymbol{\theta}_\mathrm{time}}(\widetilde{\mathbf{a}}_{n-N_\mathrm{LB}}, \ldots, \widetilde{\mathbf{a}}_{n}), \\
    \widehat{\mathbf{u}}_n &= \mathcal{H}_{\boldsymbol{\theta}_\mathrm{freq}}(\widetilde{\mathbf{a}}_{n-N_\mathrm{LB}}, \ldots, \widetilde{\mathbf{a}}_{n}),
\end{align}
where $N_\mathrm{LB} \in \mathbb{N}$ is the number of look-back frames.
We stress that both DNNs are causal and do not use magnitude at future time-frames, i.e., $N_\mathrm{LA} = 0$.

At the second stage, the phase is reconstructed from the estimated phase differences.
We design the following map $\mathcal{P}(\cdot)$ that computes the phase at the $n$th time-frame from that at the $(n-1)$th time-frame with the phase differences and magnitude:
\begin{align}
    \widehat{\boldsymbol{\phi}}_n = \mathcal{P}(\widehat{\boldsymbol{\phi}}_{n-1}, \widehat{\mathbf{v}}_n, \widehat{\mathbf{u}}_n, \mathbf{a}_n, \mathbf{a}_{n-1}),
    \label{eq:prop2}
\end{align}
where $\widehat{\mathbf{v}}_n$ is the estimated TPD computed from the estimated BPD $\widehat{\mathbf{w}}_n$.
This map is constructed based on a weighted least squares problem of complex STFT coefficients.
Its detail is postponed to Section~\ref{sec:stage2}.
Since both stages do not require information on future time-frames, the proposed framework causally reconstructs the phase in a frame-by-frame manner.

\subsection{First Stage: DNNs for Estimating Phase Differences}
\label{sec:stage1}

The proposed framework can use arbitrary causal DNNs as $\mathcal{G}_{\boldsymbol{\theta}_\mathrm{time}}(\cdot)$ and $\mathcal{H}_{\boldsymbol{\theta}_\mathrm{freq}}(\cdot)$.
While fully connected neural networks (FCNs) have been used for DNN-based phase reconstruction~\cite{Takamichi2018,Takamichi2020,Masuyama2020a,Thieling2021,Thien2021}, we present an efficient DNN architecture for the proposed framework.

According to the phase-magnitude relation, phase derivatives can be approximated by differences of log-magnitude in the surrounding T-F bins as in \eqref{eq:tdiff-ana} and \eqref{eq:fdiff-ana}.
In PGHI, the phase differences are approximated by averaging the phase derivatives at the T-F grids as in \eqref{eq:tdiff-ave-ana} and \eqref{eq:fdiff-ave-ana}.
These operations can be implemented by convolution in the T-F domain except for ${2\pi \alpha m}/{M}$ in \eqref{eq:tdiff-ana}.
In the online setting, RTPGHI uses the second order backward difference in \eqref{eq:causal}, which can also be implemented by convolution.
While these mathematical formulations are concrete, we expect that estimation accuracy can be improved by exploiting prior knowledge of a target signal.
For example, mixed derivative of phase is useful for analyzing harmonic signals~\cite{Masuyama2018}, and instantaneous frequency of a sinusoidal component can be estimated from its spectral peak~\cite{Beauregard2015}.
To acquire such complicated phase information from a dataset, DNNs should be effective.

\begin{figure}[t!]
  \centering
  \includegraphics[width=0.99\columnwidth]{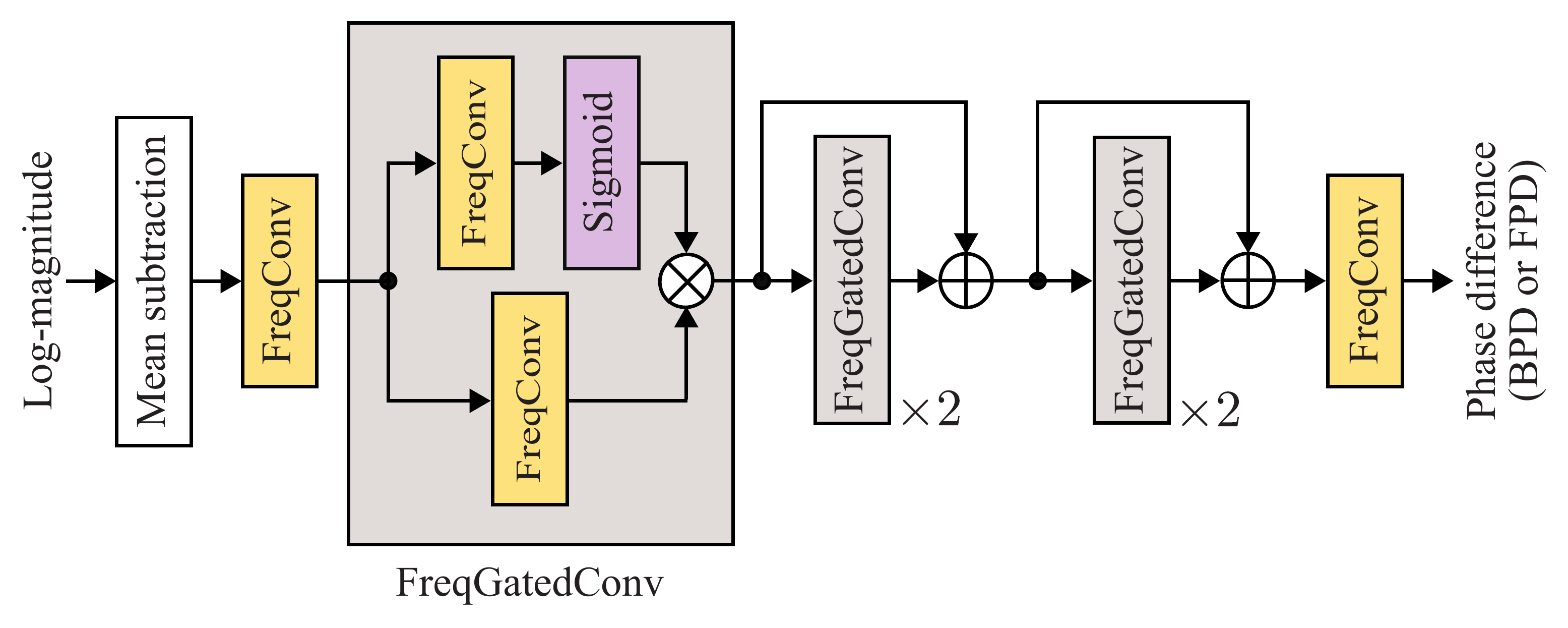}
  \caption{Illustration of a DNN for estimation of BPD or FPD.}
  \label{fig:dnn}
\end{figure}

We employ convolution layers that can efficiently aggregate information in the surrounding T-F bins.
Our DNNs consist of the mean subtraction and $1$-D frequency convolution layers (\texttt{FreqConv}) as in Fig.~\ref{fig:dnn}.
We concatenate the log-magnitude up to the current time-frame and subtract its mean:
\begin{equation}
    \boldsymbol{\Psi}_n = \mathcal{N}(\widetilde{\mathbf{a}}_{n-N_\mathrm{LB}}, \ldots, \widetilde{\mathbf{a}}_{n}), 
    \label{eq:causal-conv}
\end{equation}
where $\boldsymbol{\Psi}_n \in \mathbb{R}^{M \times (N_{\mathrm{LB}}+1)}$ is a frame-wise feature, and $\mathcal{N}(\cdot)$ subtracts the mean of its inputs.
This map just changes the global magnitude within the inputted $N_\mathrm{LB} + 1$ frames and retains the STFT phase-magnitude relation.
The following first \texttt{FreqConv} layer treats temporal adjacencies of the inputted T-F bins as channels.
As a result, the \texttt{FreqConv} layer can perform a causal convolution along the time-frame and mimic the operations in \eqref{eq:fdiff-ana} and \eqref{eq:causal}.
In detail, the number of channels of the \texttt{FreqConv} layer corresponds to the kernel size of the causal convolutions along the time-frame.
The frame-wise feature $\boldsymbol{\Psi}_n$ is passed to multiple \texttt{FreqConv} layers.
We combine the \texttt{FreqConv} layers with the gating mechanism~\cite{Dauphin2016} as follows:
\begin{align}
    &\texttt{FreqGatedConv}(\boldsymbol{\Psi}_n) \nonumber \\
    & \hspace{10pt}= \texttt{Sigmoid}(\texttt{FreqConv}(\boldsymbol{\Psi}_n)) \odot \texttt{FreqConv}( \boldsymbol{\Psi}_n), \!\!
\end{align}
where the two \texttt{FreqConv} layers have different parameters.
This mechanism can adaptively control the information passed to the next layer.
Its effectiveness has been confirmed in DNN-based phase reconstruction~\cite{Masuyama2021}.
In this first stage of the proposed framework, each feature $\boldsymbol{\Psi}_n$ is handled separately, and thus it is causal.

CNNs have difficulty of using the absolute T-F location due to their translation invariance.
In the phase-magnitude relation in \eqref{eq:tdiff-ana}, absolute frequency information $2\pi \alpha m /M$ is required to compute the phase derivative with respect to time.
It is thus difficult to estimate TPD by a CNN.
In contrast, BPD removes the absolute frequency information in \eqref{eq:bpd} and is expected to be easily estimated by a CNN.

Supervised learning is straightforward for training DNNs that estimate BPD and FPD, because the pairs of magnitude and phase differences are easily calculated from time-domain signals.
The target phase differences should be treated as circular variables because they inherit the periodic nature of the phase.
Hence, we measure the errors of the estimated phase differences by the periodic loss functions as follows:
\begin{align}
    \mathcal{L}_\mathrm{BPD}({\mathbf{W}}, \widehat{\mathbf{W}}) &=
    \sum_{m=0}^{M-1} \sum_{n=1}^{N-1} \mathcal{L}_\mathrm{cos}({W}[m,n], \widehat{W}[m,n]), \\
    \mathcal{L}_\mathrm{FPD}(\mathbf{U}, \widehat{\mathbf{U}}) &=
    \sum_{m=1}^{M-1} \sum_{n=0}^{N-1} \mathcal{L}_\mathrm{cos}(U[m,n], \widehat{U}[m,n]).
\end{align}
The range of the estimated phase differences, $\widehat{W}[m,n]$ and $\widehat{U}[m,n]$, is not restricted to $[-\pi, \pi)$.

\subsection{Second Stage: Online Phase Reconstruction From Phase Differences}
\label{sec:stage2}

\begin{figure}[t!]
  \centering
  \includegraphics[width=0.9\columnwidth]{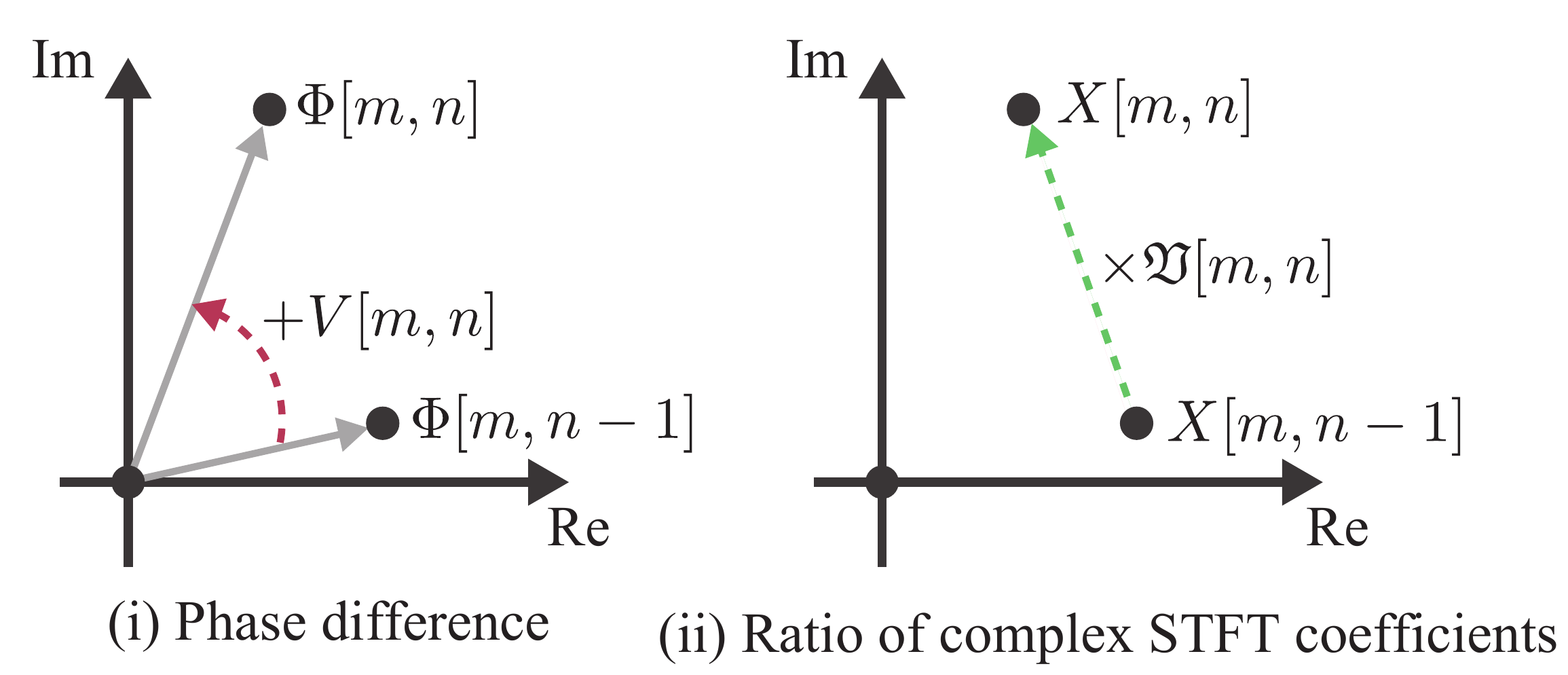}
  \caption{Illustration of the conversion from (\romaone) the phase difference to (\romatwo) the ratio of the complex STFT coefficients.}
  \label{fig:ang2comp}
\end{figure}

As in \eqref{eq:prop2}, the second stage of the proposed framework $\mathcal{P}(\cdot)$ recurrently estimates the phase based on the estimated phase differences.
The phase differences estimated by DNNs have ambiguity of $2\pi$ due to the use of the periodic loss function.
The second stage of the proposed framework must take care of this ambiguity.
Since it is not easy to directly handle such phase differences with the ambiguity, we propose to convert them to the ratios of complex STFT coefficients as follows:
\begin{align}
    \widehat{\mathfrak{V}}[m,n] &= \frac{A[m,n]}{A[m,n-1]}\mathrm{e}^{\mathrm{i} \widehat{V}[m,n]}, \label{eq:ifconvert} \\
    \widehat{\mathfrak{U}}[m,n] &= \frac{A[m,n]}{A[m-1,n]}\mathrm{e}^{\mathrm{i} \widehat{U}[m,n]}, \label{eq:gdconvert}
\end{align}
where $\widehat{V}[m,n] = \widehat{W}[m,n] + {2\pi \alpha m}/{M}$.
Note that the oracle versions of these ratios are given by
\begin{align}
    {\mathfrak{V}}[m,n] &= \frac{X[m,n]}{X[m,n-1]}, \\
    {\mathfrak{U}}[m,n] &= \frac{X[m,n]}{X[m-1,n]},
\end{align}
which cannot be computed because the phase of $X[m,n]$ is not available. 
This conversion is depicted in Fig.~\ref{fig:ang2comp}.
The main advantage of this conversion is that the $2\pi$ ambiguity of the estimated phase differences is avoided.

On the basis of the complex ratios, we formulate an optimization problem and estimate the phase by solving it.
Let us consider the complex ratios at the $n$th time-frame $\widehat{\boldsymbol{\mathfrak{v}}}_n$ and $\widehat{\boldsymbol{\mathfrak{u}}}_n$ given by
\begin{align}
\widehat{\boldsymbol{\mathfrak{v}}}_n &= [\widehat{\mathfrak{V}}[0,n], \ldots, \widehat{\mathfrak{V}}[M-1,n]]^\mathsf{T}, \\ \widehat{\boldsymbol{\mathfrak{u}}}_n &= [\widehat{\mathfrak{U}}[1,n], \ldots, \widehat{\mathfrak{U}}[M-1,n]]^\mathsf{T}.
\end{align}
To enforce the complex ratios between successive time-frames close to $\widehat{\boldsymbol{\mathfrak{v}}}_n$, we minimize the following function $\mathscr{T}(\cdot)$ with respect to an optimization variable $\mathbf{z}_n \in \mathbb{C}^M$:
\begin{equation}
    \mathscr{T}(\mathbf{z}_n, \widehat{\mathbf{x}}_{n-1}, \widehat{\boldsymbol{\mathfrak{v}}}_n)
    = \| \mathbf{z}_n - \mathrm{diag}(\widehat{\mathfrak{v}}_n) \widehat{\mathbf{x}}_{n-1} \|_{\boldsymbol{\Lambda}_n}^2, \label{eq:ifloss}
\end{equation}
where $\widehat{\mathbf{x}}_{n-1} = [\widehat{X}[0,n-1], \ldots,\widehat{X}[M-1,n-1]]^\mathsf{T}$ is the estimated STFT coefficients at the previous time-frame,
$\mathrm{diag}(\cdot)$ returns the diagonal matrix whose diagonal elements are its input vector, and $\| \mathbf{z} \|_{\boldsymbol{\Lambda}_n}^2 = \mathbf{z}^\mathsf{H} {\boldsymbol{\Lambda}_n} \mathbf{z}$.
Meanwhile, to enforce the complex ratios between adjacent frequencies close to $\widehat{\boldsymbol{\mathfrak{u}}}_n$, the following function $\mathscr{S}(\cdot)$ is also minimized:
\begin{equation}
    \mathscr{S}(\mathbf{z}_n, \widehat{\boldsymbol{\mathfrak{u}}}_n )
    = \| \mathbf{D}_n \mathbf{z}_n \|_{\boldsymbol{\Gamma}_n}^2,
    \label{eq:gdloss}
\end{equation}
where the matrix $\mathbf{D}_n \in \mathbb{R}^{M-1 \times M}$ is defined as
\begin{align}
    D_n[m-1,m-1] &= - \widehat{\mathfrak{U}}[m,n], \\
    D_n[m-1,m] &= 1,
\end{align}
and the other entries are zero.
The weights $\boldsymbol{\Lambda}_n$ in \eqref{eq:ifloss} and $\boldsymbol{\Gamma}_n$ in \eqref{eq:gdloss} are diagonal matrices that reflect the reliability of the estimated TPD and FPD, respectively.
The detail of the weights is explained in the next subsection.
By using these two functions, the map $\mathcal{P}(\cdot)$ in \eqref{eq:prop2} is realized as follows:
\begin{align}
\widehat{\boldsymbol{\phi}}_n &= \boldsymbol{\mathtt{Arg}}(\undertilde{\mathbf{x}}_n), \label{eq:2nd-stage} \\
\undertilde{\mathbf{x}}_n &=
\argmin_{\mathbf{z}_n}
\mathscr{T}(\mathbf{z}_n, \widehat{\mathbf{x}}_{n-1}, \widehat{\boldsymbol{\mathfrak{v}}}_n) + \mathscr{S}(\mathbf{z}_n, \widehat{\boldsymbol{\mathfrak{u}}}_n ), \label{eq:objective-functions}
\end{align}
where \eqref{eq:2nd-stage} calculates the complex-argument element-wise, i.e., $\widehat{\Phi}[m,n] = \mathrm{Arg}(\undertilde{X}[m,n])$.
The solution of \eqref{eq:objective-functions} $\undertilde{\mathbf{x}}_n = [\undertilde{X}[0,n], \ldots, \undertilde{X}[M-1,n]]^\mathsf{T}$ does not maintain the given magnitude $A[m,n]$.
We thus modify it as $\widehat{X}[m,n] = A[m,n] \exp(\mathrm{i} \widehat{\Phi}[m,n])$ and use the modified version in \eqref{eq:ifloss} for the next time-frame.

The optimization problem in \eqref{eq:objective-functions} aims to estimate complex STFT coefficients that are consistent with the ratios calculated from the phase differences.
It can be solved in a closed form:
\begin{equation}
    \undertilde{\mathbf{x}}_n =  (\boldsymbol{\Lambda}_n + \mathbf{D}_n^\mathsf{T} \boldsymbol{\Gamma}_n \mathbf{D}_n)^{-1} \boldsymbol{\Lambda}_n \mathbf{y}_n, \label{eq:crpusolve}
\end{equation}
where the $m$th entry of $\mathbf{y}_n$ is given by $\widehat{\mathfrak{v}}[m,n] \widehat{X}[m,n-1]$. 
By using \eqref{eq:crpusolve}, the proposed method reconstructs the phase in a frame-by-frame manner without any iterative optimization.

\subsection{Weighting Rule and Initialization}
\label{sec:weight-rule}

In the optimization-based phase reconstruction given in \eqref{eq:2nd-stage} and \eqref{eq:objective-functions}, the weights, $\boldsymbol{\Lambda}_n$ and $\boldsymbol{\Gamma}_n$, are important to improve the quality of the reconstructed signal.
The phase differences with large weights are maintained, and thus the weights must be designed based on the reliability of the estimated phase differences.
We propose to design the $m$th diagonal entry of $\boldsymbol{\Lambda}_n$ and $\boldsymbol{\Gamma}_n$ by using the given magnitude:
\begin{align}
    \Lambda_n[m,m] &= (A[m,n] A[m,n-1])^{p},
    \label{eq:weight-time} \\
    \Gamma_n[m,m] &= \gamma_0 (A[m,n] A[m-1,n])^{p},
    \label{eq:weight-freq}
\end{align}
where $p$ is a parameter for compressing or enhancing the magnitudes, and $\gamma_0 \geq 0$ is a parameter to balance the two weights.
These weights are based on the assumption that the ratios of complex STFT coefficients are accurate when magnitude at the related T-F bins is large.

As a special case, the proposed method in \eqref{eq:crpusolve} results in the integration of the estimated TPD over time when $\gamma_0 = 0$:
\begin{equation}
    \widehat{\Phi}[m,n] = \widehat{\Phi}[m,n-1] + \widehat{V}[m,n]. \label{eq:ifint}
\end{equation}
This phase reconstruction was already used in a DNN-based method~\cite{Engel2019}.
Its performance is limited because the relation between adjacent STFT coefficients in the frequency direction is neglected.
The proposed method with $\gamma_0 > 0$ uses the relations in both time and frequency directions.
Another related work \cite{Nugraha2019} applies some weight designed from the given magnitude to training of DNNs that estimate phase.
In contrast, we use the weights for reconstructing phase from the phase differences but not for training DNNs.

The recurrent phase reconstruction in \eqref{eq:crpusolve} is not applicable to the initial time-frame because $\widehat{\mathbf{x}}_{n-1}$ is not given.
We compute the phase at the initial time-frame $\widehat{\Phi}[m,0]$ by accumulating the estimated FPD.
In our preliminary experiments, however, it often resulted in a similar performance with other initialization methods, e.g., the zero and random phases.
This should be because the estimated FPD is unreliable due to a small magnitude at the initial time-frame.
The proposed method is robust against errors at the T-F bins with small magnitudes because the relations between the T-F bins with large magnitudes are emphasized by the weights in \eqref{eq:weight-time}--\eqref{eq:weight-freq}.
In detail, if the T-F bins at the previous time-frame have small magnitude, the proposed method tries to maintain the estimated FPD at the current time-frame and neglect the phase at the previous time-frame.

\section{Experiments}
\label{sec:exp}

In this section, we investigate the performance of the proposed DNN-based two-stage framework in online phase reconstruction.
The experimental conditions are described in Section~\ref{sec:condition}.
Section~\ref{sec:exp-performance} compares the proposed framework with various online and offline phase reconstruction methods.
The generalization capability and robustness of the proposed framework are shown in Sections~\ref{sec:exp-vctk} and \ref{sec:exp-mel}, respectively.
The effectiveness of our CNN for the first stage is validated in Section~\ref{sec:exp-1st-stage}.
We investigate the effect of the weight parameters, $p$ and $\gamma_0$, on the quality of the reconstructed signals in Section~\ref{sec:exp-2nd-stage-1}.
Section~\ref{sec:exp-2nd-stage-2} demonstrates the effectiveness of the optimization-based phase reconstruction method in the second stage.
Finally, the proposed two-stage framework is compared with direct phase reconstruction in Section~\ref{sec:exp-vs-direct}.

\subsection{Experimental Conditions}
\label{sec:condition}

\subsubsection{Dataset and STFT Parameters}
\label{sec:5a1}

Evaluations were performed on the LJ speech dataset that consists of $13100$ audio clips uttered by a female speaker.
The audio clips were sampled at $22050$ Hz and randomly splitted into three subsets: $12500$ clips for training, $300$ clips for validation, and $300$ clips for testing as in \cite{Prenger2019}.
During the training, the utterances were further divided into about $1$-second-long segments ($24064$ samples).
The validation set was used to optimize the hyperparameters for the second stage.
STFT was computed with the Hann window, where the window size and shift size were $1024$ and $256$ samples, respectively.
We used \texttt{ltfatpy}%
\footnote{
\texttt{ltfatpy} is available under: \url{https://dev.pages.lis-lab.fr/ltfatpy/}.
It is a python version of \texttt{LTFAT}: \url{http://ltfat.org/}~\cite{Prusa2013}.
}
to implement STFT and related transformations.

\subsubsection{DNN Configuration and Training Setup}
\label{sec:5a2}

The DNN used in the following experiments is illustrated in Fig.~\ref{fig:dnn}.
It consists of the mean subtraction layer, a \texttt{FreqConv} layer, and five \texttt{FreqGatedConv} layers followed by another \texttt{FreqConv} layer.
We set the number of look-back frames $N_\mathrm{LB}$ to $3$ based on the overlap of the window for STFT.
Other configurations are summarized in Table~\ref{tab:condition}.

To train the DNNs, we used the RAdam optimizer~\cite{Liu2020} for $100$ epoch where the batch size was $32$.
We linearly warmed up the learning rate for $5$ epochs to $0.0004$ and adopt the half-period cosine scheduler~\cite{Loshchilov2017}.
We applied a weight decay of $10^{-6}$ and a gradient clipping of $10$ for stable training, which were implemented in \texttt{Pytorch}~\cite{Paszke2019}.

{
\begin{table}[t!]
\centering
\caption{Experimental Conditions}
\label{tab:condition}
\begin{tabular}{c|c}
\hline
\hline
\multicolumn{2}{c}{Parameters of DNN Architecture} \\
\hline
$\#$ of \texttt{FreqConv} layers & $1+1$ \\
$\#$ of \texttt{FreqGatedConv} layers & $5$ \\
$\#$ of channels & $64$ \\
Kernel size of \texttt{FreqGatedConv} layers & $3$ \\
Kernel size of \texttt{FreqConv} layers & $1$ \\
$N_{\mathrm{LB}}$ & $3$ \\
$\#$ of parameters & $206$k \\
\hline
\multicolumn{2}{c}{Parameters for Training} \\
\hline
Optimizer & RAdam~\cite{Liu2020} \\
Base learning rate & $0.0004$ \\
Batch size & $32$ \\
$\#$ of epochs & $100$ \\
$\#$ of warmup epochs & $5$ \\
Weight decay & $10^{-6}$ \\
Maximum norm of the gradients & $10$ \\
\hline \hline
\end{tabular}
\end{table}
}

\subsubsection{Evaluation Metrics}

The results of phase reconstruction were evaluated by three objective measures.
The first one is the log-spectral convergence (LSC)~\cite{Strumel2011} defined by
\begin{equation}
    \text{LSC}(\widehat{\mathbf{X}}, \mathbf{A}) \!= 20\log_{10}\frac{\| \mathbf{A} - |\mathrm{STFT}(\mathrm{iSTFT}(\widehat{\mathbf{X}}))| \|_\mathrm{Fro}}{\| \mathbf{A} \|_\mathrm{Fro}},
    \!\!
\end{equation}
where $\widehat{X}[m,n] = A[m,n] \, \mathrm{e}^{\widehat{\Phi}[m,n]}$, and $\| \cdot \|_{\mathrm{Fro}}$ denotes the Frobenius norm.
When the estimated phase is perfect, i.e., $\widehat{\boldsymbol{\Phi}} = \boldsymbol{\Phi}$, iSTFT and STFT do not alter the magnitude of $\widehat{\mathbf{X}}$, and LSC becomes $-\infty$.
The second and third ones are the wide-band extension of the perceptual evaluation of subjective quality (PESQ)~\cite{wpesq} and the extended short-time objective intelligibility (ESTOI)~\cite{Jensen2016}.
These objective measures have been commonly used to evaluate naturalness and intelligibility of the results of phase-aware speech enhancement~\cite{Zheng2019} and separation~\cite{Wang2021}.

\begin{figure}[t!]
  \centering
  \includegraphics[width=0.99\columnwidth]{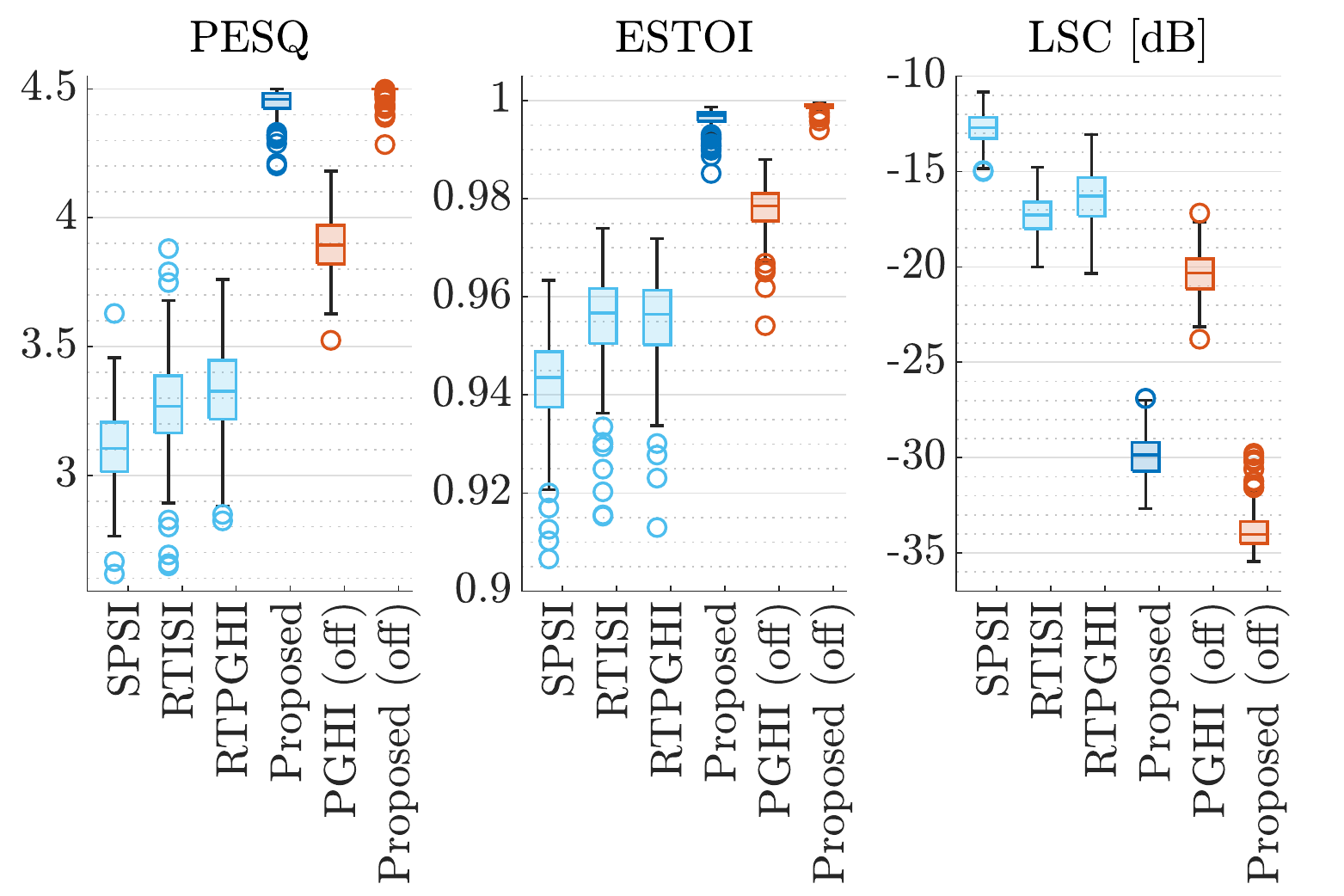}
  \caption{Boxplots of PESQ, ESTOI, and LSC for $300$ reconstructed utterances.
  Blue and red boxes correspond to online and offline phase reconstruction methods, respectively.
  Higher PESQ and ESTOI indicate better sound quality.
  Lower LSC indicates better phase reconstruction.
  }
  \label{fig:comparison}
\end{figure}

\subsection{Comparison to Existing Online Phase Reconstruction}
\label{sec:exp-performance}

To validate the effectiveness of the proposed DNN-based phase reconstruction, we compared the proposed method with three online phase reconstruction methods: RTPGHI~\cite{Prusa2016}, a consistency-based phase reconstruction method called real-time iterative spectrogram inversion (RTISI)~\cite{Zhu2007}, and a sinusoidal-model-based phase reconstruction method called single pass spectrogram inversion (SPSI)~\cite{Beauregard2015}.
For RTPGHI and RTISI, we set the number of look-ahead frames $N_\mathrm{LA}$ to $0$, i.e., all methods were set causal.
We also investigated the performance of two offline methods: PGHI and an offline version of the proposed method.
In the offline proposed method, we additionally concatenated three look-ahead frames as the input of the DNNs.
All existing methods are implemented in \texttt{PHASERET}~\cite{Prusa2017a}.
In RTISI, the number of per-time-frame iterations was set to $5$, which is the default value of \texttt{PHASERET}.
We would like to stress that this is not a fair comparison because only the proposed method uses a DNN trained on utterances of the target speaker.
This comparison, however, demonstrates the great potential of incorporating the prior knowledge into online phase reconstruction.
The generalization capability of the proposed method is validated in the next subsection.
Furthermore, the performance of the existing methods combined with the DNNs are investigated in Section~\ref{sec:exp-2nd-stage-2}.

PESQ, ESTOI, and LSC of the reconstructed signals are summarized in Fig.~\ref{fig:comparison}.
The two methods that do not utilize prior knowledge of target signals, RTPGHI and RTISI, resulted in similar performance.
Although SPSI considers the sinusoidal model for target signals, it performed worse than the other methods in our experiment.
This should be a consequence of the mismatch between the signal model and actual signals.
The proposed method was able to outperform all of the existing methods.
Note that the proposed method also outperformed the offline PGHI.
This result confirms the advantage of leveraging prior knowledge of the target signals learned by DNNs.
The offline proposed method substantially improved the performance from that of the online version.
That is, the performance of the two-stage phase reconstruction can be improved by leveraging look-ahead frames.

\subsection{Generalization for Unseen Speakers}
\label{sec:exp-vctk}

To clarify the generalization capability of the proposed method, we evaluated it on unseen speakers.
While the training and validation data were from the LJ speech dataset, the evaluation was performed on utterances of three females (\texttt{p225}, \texttt{p228}, \texttt{p229}) and three males (\texttt{p226}, \texttt{p227}, \texttt{p232}) from the VCTK corpus.
For each speaker, $100$ utterances were randomly selected and resampled at $22050$ Hz as in \cite{Masuyama2021}.

The experimental results are summarized in Table~\ref{tab:vctk}.
Even on the unseen speakers, the proposed method outperformed RTPGHI which was the best reference method in the previous experiment.
This result confirms the generalization capability of the proposed framework even when the training dataset consists of utterances of a single speaker.
Training on utterances of multiple speakers might improve the generalization capability further.

\begin{table}[t]
    \centering
    \caption{Median of PESQ and ESTOI of $100$ reconstructed utterances for each speaker.}
    \scalebox{1.}[1.]{
    \begin{tabular}{c|ccc|ccc}
        \toprule
        & \multicolumn{3}{c|}{Female} & \multicolumn{3}{c}{Male} \\
        \cmidrule{2-7}
        & \texttt{p225} & \texttt{p228} & \texttt{p229} & \texttt{p226} & \texttt{p227} & \texttt{p232} \\
        \midrule
        \multicolumn{1}{c}{} & \multicolumn{6}{c}{PESQ} \\
        \midrule
        RTPGHI & 3.52 & 3.35 & 3.27 & 3.40 & 3.19 & 3.38 \\
        Proposed & \bf{4.31} & \bf{4.22} & \bf{4.25} & \bf{4.27} & \bf{4.29} & \bf{4.31} \\
        \midrule
        \multicolumn{1}{c}{} & \multicolumn{6}{c}{ESTOI} \\
        \midrule
        RTPGHI & 0.840 & 0.830 & 0.859 & 0.823 & 0.862 & 0.874 \\
        Proposed & \bf{0.949} & \bf{0.959} & \bf{0.969} & \bf{0.957} & \bf{0.969} & \bf{0.975} \\
        \bottomrule
    \end{tabular}
    }
    \label{tab:vctk}
\end{table}

\subsection{Application to Mel-Spectrogram Inversion}
\label{sec:exp-mel}

In many applications, the given STFT magnitude contains some errors.
To investigate the robustness against such errors, we validated RTPGHI and the proposed method on the magnitude recovered from that compressed to the mel scale.
Mel-spectrograms have been widely used as acoustic features in audio synthesis, and phase reconstruction has been applied to the magnitudes recovered from them~\cite{Lee2019,Yang2019}.
In this experiment, the power-compressed magnitude in the linear scale with $513$ bins, $A[m,n]^{0.3}$, was converted to the mel scale with a smaller number of bins $M_\text{mel} \in \{ 80, 160, 240, 320, 400 \}$.
Then, the mel-scale magnitude is converted back to that in the linear scale with regular power by solving a nonnegative least squares problem, which is implemented in \texttt{Librosa}~\cite{McFee2015}.
The power compression has been used to maintain the components with a small magnitude in least squares as in~\cite{Wisdom2019} and was effective for both RTPGHI and the proposed method.
The smaller $M_\text{mel}$ caused more error in the recovered linear-scale magnitude.

\begin{figure}[t!]
  \centering
  \includegraphics[width=0.99\columnwidth]{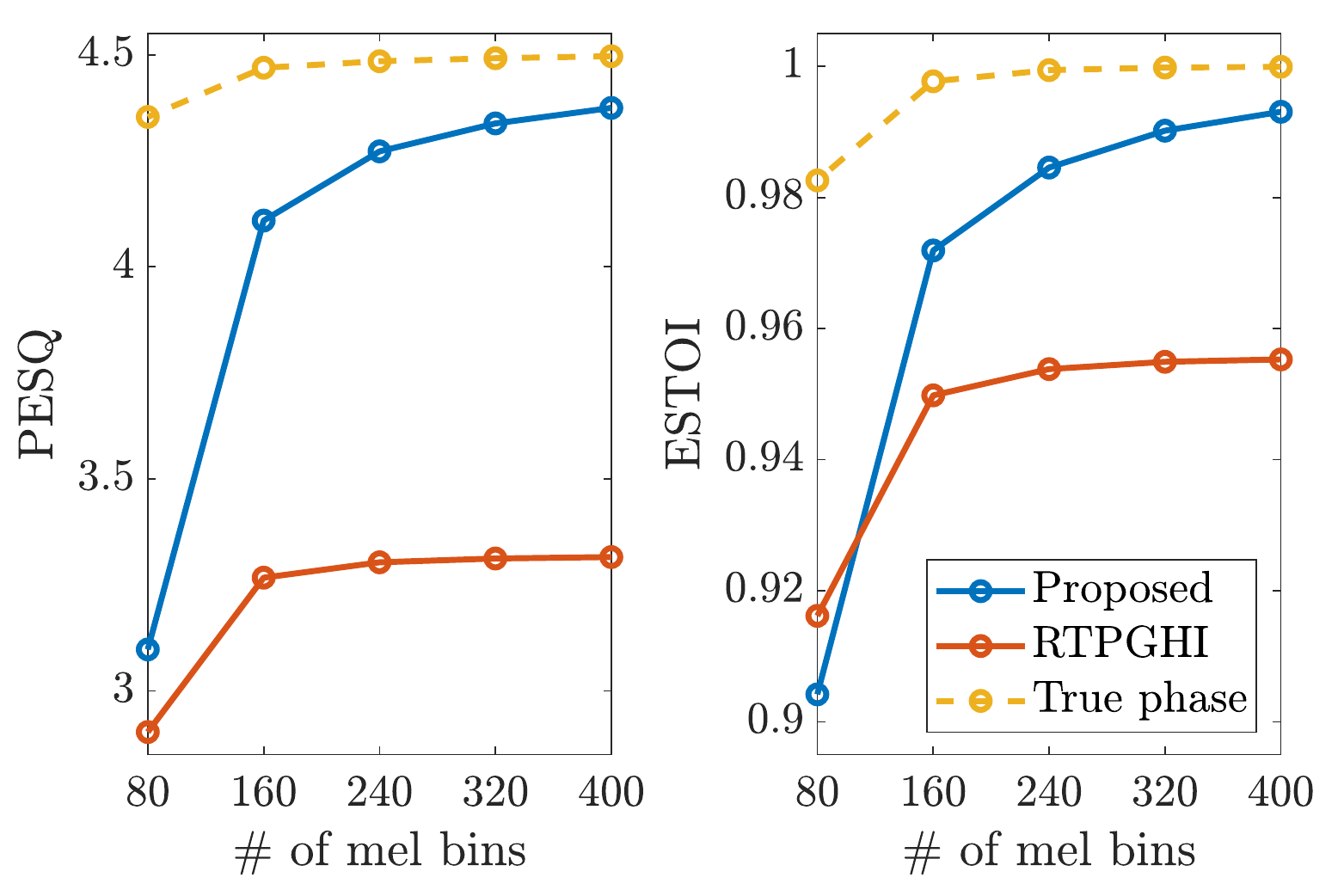}
  \caption{Average PESQ and ESTOI of utterances reconstructed from the degraded STFT magnitudes.
  The magnitudes in the linear scale with $513$ bins were recovered from the mel-spectrograms in different number of bins.}
  \label{fig:mel}
\end{figure}

PESQ and ESTOI of the reconstructed signals are shown in Fig.~\ref{fig:mel}.
In addition to RTPGHI and the proposed method, we evaluated the recovered magnitude with the true phase.
This is an upper bound of the performance of phase reconstruction.
When $M_\text{mel} > 80$, the proposed method substantially outperformed RTPGHI.
Recently, DNNs have been used to recover the magnitude in the linear scale from that in the mel scale~\cite{Lee2019,Yang2019}.
These DNN-based methods should improve the quality of the recovered magnitude from the nonnegative least squares used in this experiment.
We thus expect that the performance of the proposed method is improved by incorporating it with the DNN-based estimation of the magnitude in the linear scale.

\begin{figure}[t!]
  \centering
  \includegraphics[width=0.99\columnwidth]{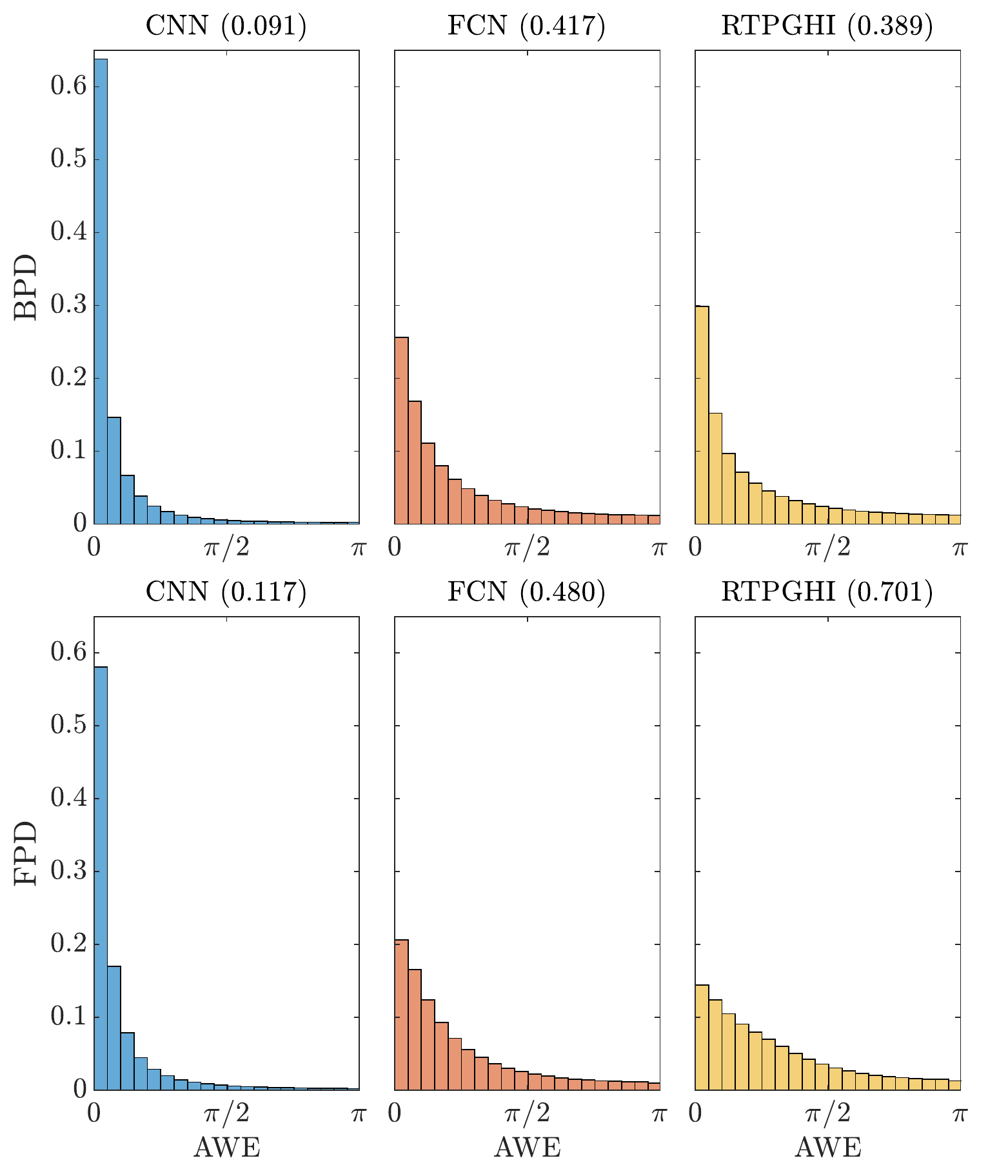}
  \caption{Histogram of the absolute wrapped error (AWE) of estimated BPD and FPD.
  The vertical axis is proportion to the total number of all T-F bins and audio clips.
  The number inside the above parentheses represents median.
  }
  \label{fig:bpd-fpd}
\end{figure}

\subsection{Effectiveness of CNN to Estimate BPD and FPD}
\label{sec:exp-1st-stage}

To validate the effectiveness of the proposed CNN for estimating BPD and FPD, we compared its estimation accuracy with that of an FCN.
The CNN was the same as that used in the previous experiment (Fig.~\ref{fig:dnn}).
The FCN comprised $3$ gated linear units of $1024$ units and a linear output layer as in \cite{Takamichi2018}, where its number of total parameters was $8929$k which is about $43$ times more than that of the CNN.
The input of the FCN was the log-magnitudes up to the current time-frame as in \eqref{eq:causal-conv}, but we concatenated them along with the frequency direction.
The training configuration was the same as in Section~\ref{sec:5a2}.
The estimated phase differences were evaluated by the following absolute wrapped error (AWE):
\begin{equation}
    \mathcal{L}_\text{abs}(\phi, \widehat{\phi}) = \left| \mathcal{W}(\phi -  \widehat{\phi}) \right|.
    \label{eq:wrapped-error}
\end{equation}

The histograms of AWE of the estimated BPD and FPD are illustrated in Fig.~\ref{fig:bpd-fpd}.
These histograms are more biased towards the left when the estimates were more accurate.
AWE of RTPGHI is summarized in the rightmost column, where TPD computed by \eqref{eq:tdiff-ave-ana} was converted to BPD.
The accuracy of FPD was significantly worse than that of BPD because RTPGHI in the causal setting must use the second order backward difference in \eqref{eq:causal} instead of the centered difference for computing FPD.
If the second order centered difference was used by allowing one look-ahead frame, the median of AWE was reduced to $0.389$ from $0.701$.
Even though the DNNs did not use any look-ahead frames, they achieved notably better accuracy compared to RTPGHI.
By efficiently aggregating information in the surrounding T-F bins, the CNN outperformed the FCN with $43$ times fewer parameters.
Consequently, as shown in Table~\ref{tab:cnn-fcn}, the objective measures of the reconstructed utterances were significantly improved by using the CNN for the estimation of phase differences.

\begin{figure}[t!]
  \centering
  \includegraphics[width=0.99\columnwidth]{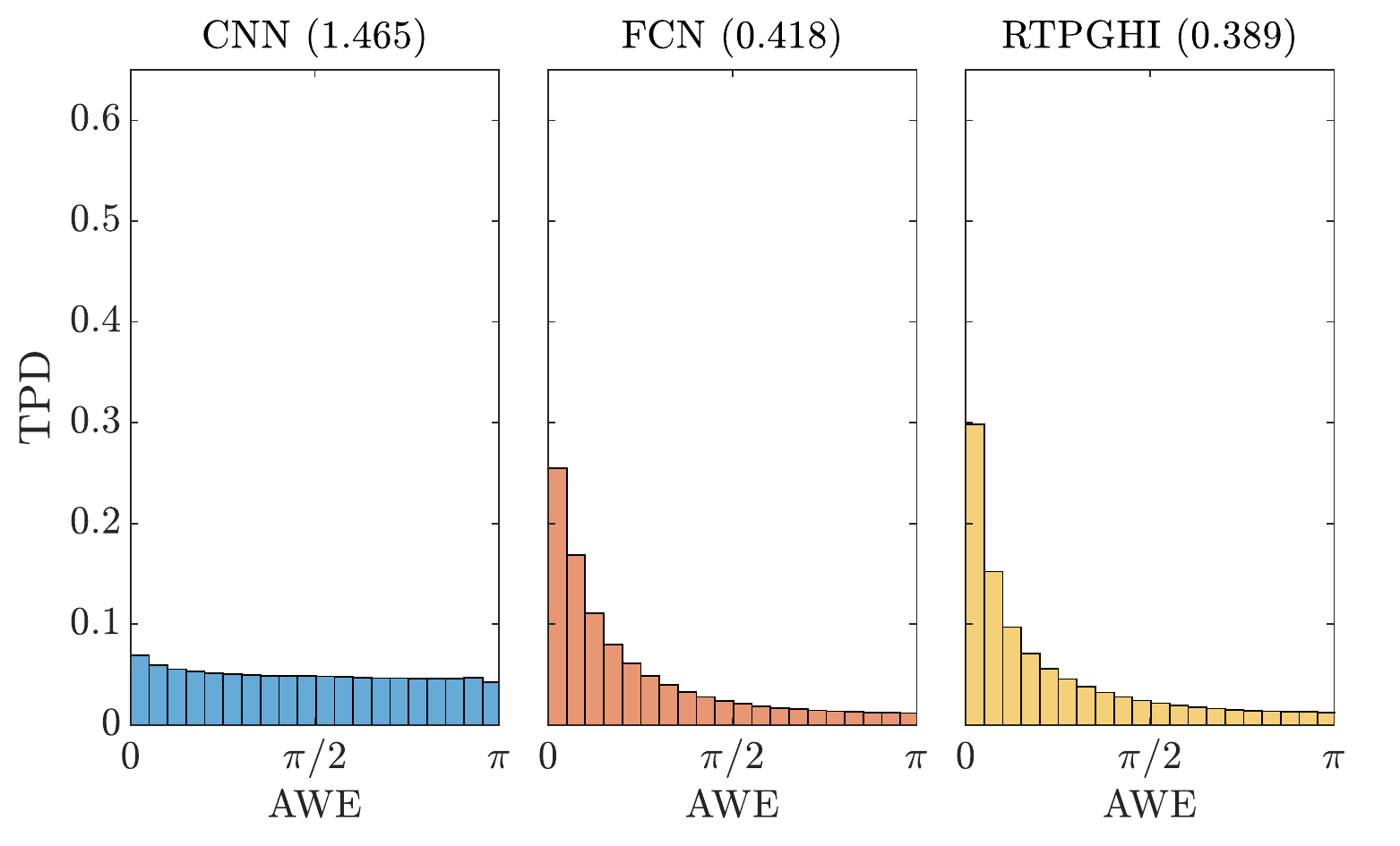}
  \caption{Histogram of AWE of estimated TPD.}
  \label{fig:tpd}
\end{figure}

To demonstrate the difficulty of estimating TPD using a CNN, we compared a CNN and an FCN by directly estimating TPD.
Note that the proposed framework does not estimate TPD itself, and hence we trained another DNN for this experiment.
The histograms of AWE for TPD estimation are depicted in Fig.~\ref{fig:tpd}.
The FCN achieved performance similar to that for BPD in Fig.~\ref{fig:bpd-fpd}.
In contrast, TPD estimation by CNN resulted in much more error compared to that of BPD.
This result indicates that estimation of TPD is difficult for the CNN as discussed in Section~\ref{sec:stage1}.
Hence, the conversion of the target from TPD to BPD is essential for the CNN.

\begin{table}[t]
    \centering
    \caption{Median of PESQ and ESTOI of utterances reconstructed from the phase differences estimated by CNN and FCN.}
    \scalebox{1.}[1.]{
    \begin{tabular}{c|ccc}
    \toprule
    & PESQ & ESTOI & LSC [dB]\\
    \midrule
    CNN & \bf{4.46} & \bf{0.997} & \bf{-29.87} \\
    FCN & 4.12 & 0.983 & -18.64 \\
    \bottomrule
    \end{tabular}
    }
    \label{tab:cnn-fcn}
\end{table}

\subsection{Effect of the Weight Parameters $p$ and $\gamma_0$}
\label{sec:exp-2nd-stage-1}

As discussed in Section~\ref{sec:weight-rule}, the weights are important in the second stage of the proposed framework.
In our experiments, the optimal weight parameters were obtained by using the validation set, where the phase differences were estimated by the CNNs trained as Section~\ref{sec:condition}.
The search range of $p$ and $\gamma_0$ were set to $[0.1, 10]$ and $[1, 100]$, respectively.

\begin{figure}[t!]
  \centering
  \includegraphics[width=0.99\columnwidth]{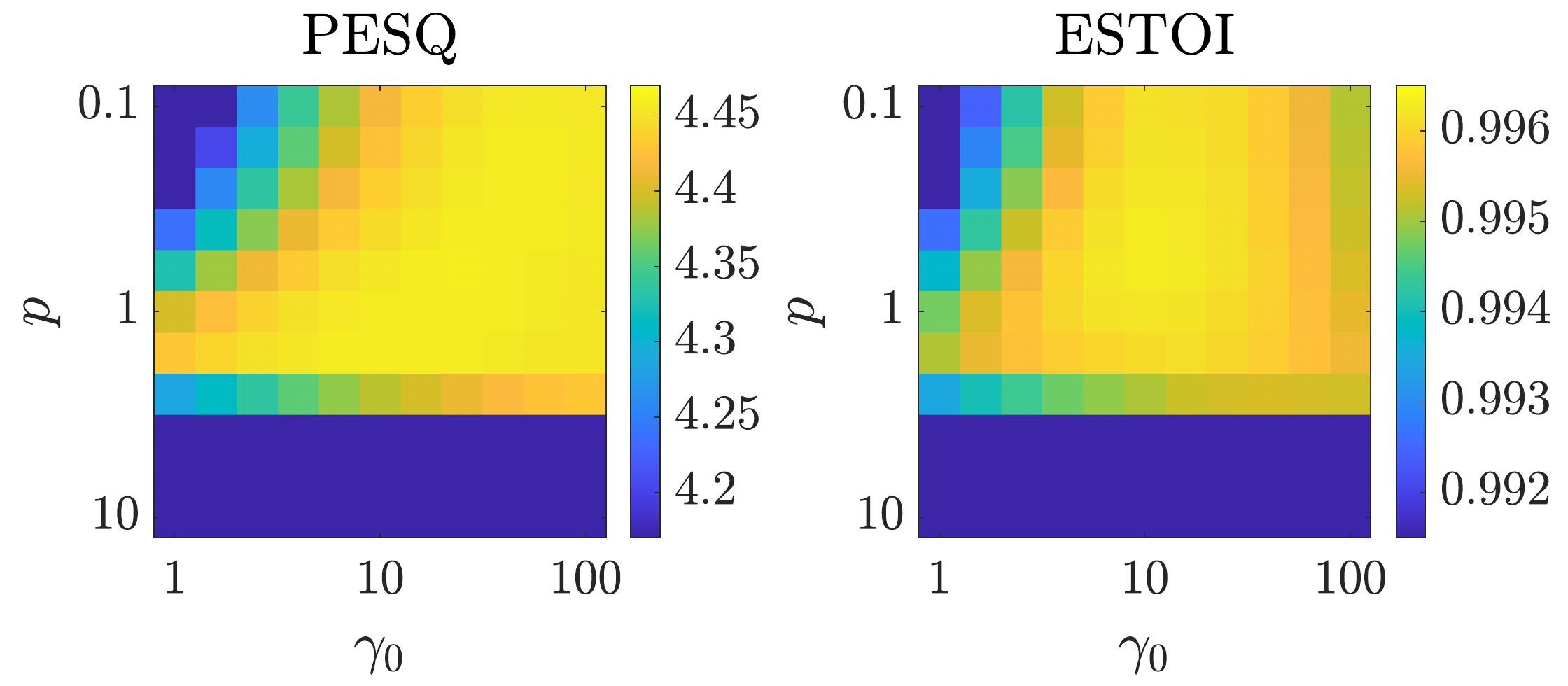}
  \caption{Average PESQ and ESTOI on the validation set.
  The parameters varied on the logarithmic scale in both axes.
  We limit the color ranges to clarify the peaks, which results in saturation for $p > 10^{0.4}$.
  }
  \label{fig:ave-pesq-estoi-validation}
\end{figure}

Fig.~\ref{fig:ave-pesq-estoi-validation} shows PESQ and ESTOI of the reconstructed signals on the validation set.
The proposed framework performed well with wide range of parameters, i.e., it is not so sensitive to these parameters.
ESTOI took the maximum value $0.996$ at $p=10^{-0.4}$ and $\gamma_0 = 10$, which also gives a high PESQ value.
Hence, these parameters were used in the other experiments.

\subsection{Evaluation of Various Methods for Reconstructing Phase From Estimated Phase Differences}
\label{sec:exp-2nd-stage-2}

In this experiment, the second stage of the proposed framework was evaluated.
We compared the second stage of the proposed framework with existing methods:
the time integration of the estimated TPD used in \cite{Engel2019} defined by \eqref{eq:ifint}, and the recurrent phase unwrapping (RPU) presented in our conference paper~\cite{Masuyama2020a}.
The difference between RPU and the proposal of this paper is the definition of least squares problems; RPU solves the least squares problem of phase without weighting, while the second stage of the proposed framework solves the least squares problem of complex STFT coefficients with weighting.
We also evaluated the adaptive integration of TPD and FPD used in RTPGHI~\cite{Prusa2016}.
Note that, if the oracle phase differences are available, all methods can perfectly reconstruct the phase up to the global constant.
To investigate their robustness to the error in the phase differences, this experiment used the analytic formulas in \eqref{eq:tdiff-ave-ana} and \eqref{eq:fdiff-ave-ana} or the CNNs to estimate the phase differences.

\begin{figure}[t!]
  \centering
  \includegraphics[width=0.99\columnwidth]{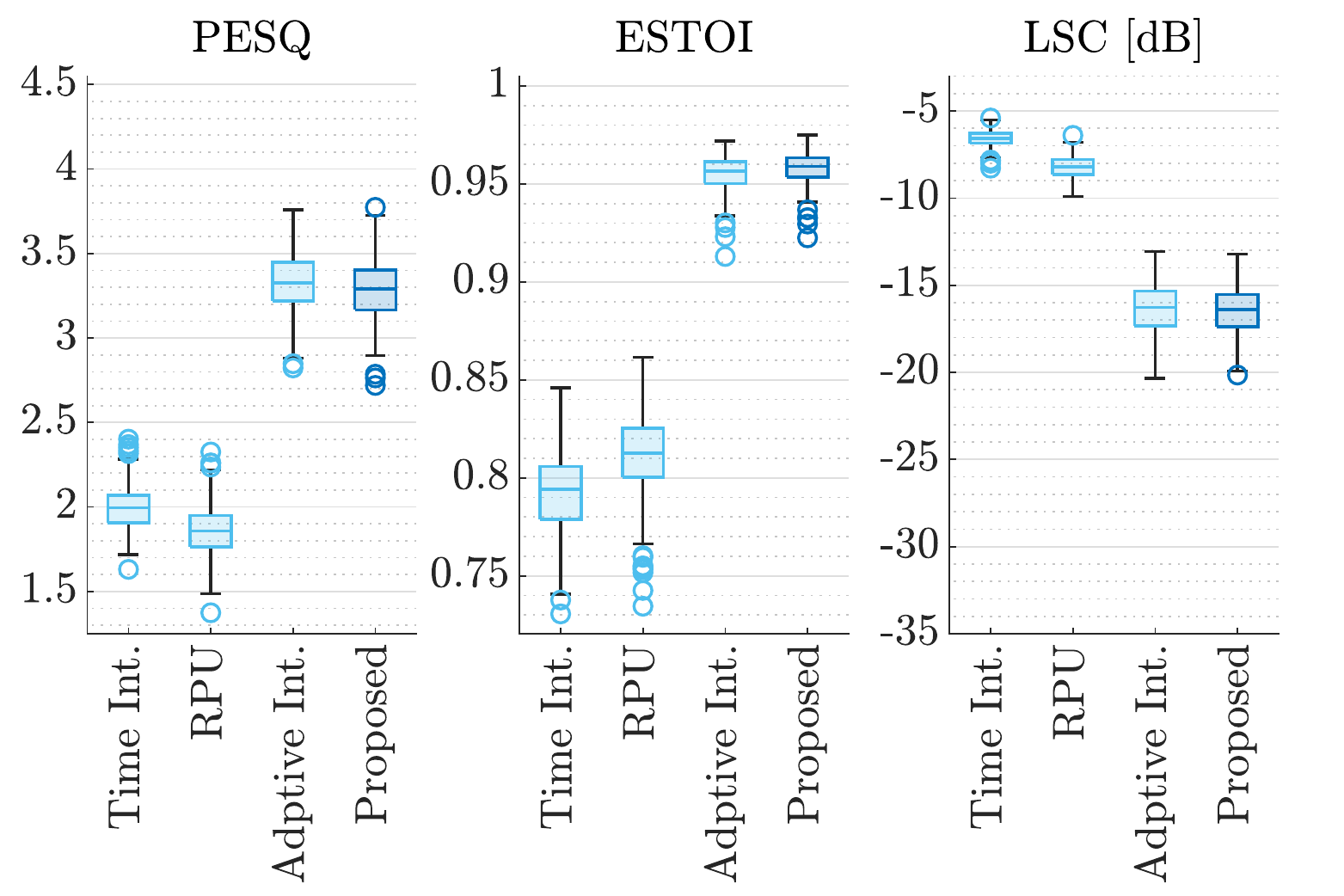}
  \caption{Boxplots of PESQ, ESTOI, and LSC of signals reconstructed from the phase differences approximated by \eqref{eq:tdiff-ave-ana} and \eqref{eq:fdiff-ave-ana}.
  The time integration of TPD \cite{Engel2019} and the adaptive integration of TPD and FPD \cite{Prusa2016} are abbreviated as Time Int.\ and Adaptive Int., respectively.
  }
  \label{fig:exp-2nd-stage-rtpghi}
\end{figure}

Fig.~\ref{fig:exp-2nd-stage-rtpghi} summarizes the results using the phase differences computed by the analytic formulas in \eqref{eq:tdiff-ave-ana} and \eqref{eq:fdiff-ave-ana}.
The performance of the adaptive integration and the proposed method was significantly better than that of the time integration and RPU.
The former group uses the magnitude to reconstruct the phase from the estimated phase differences, while the latter group does not.
As a result, the former group is more robust to the estimation error at T-F bins with small magnitudes.

\begin{figure}[t!]
  \centering
  \includegraphics[width=0.99\columnwidth]{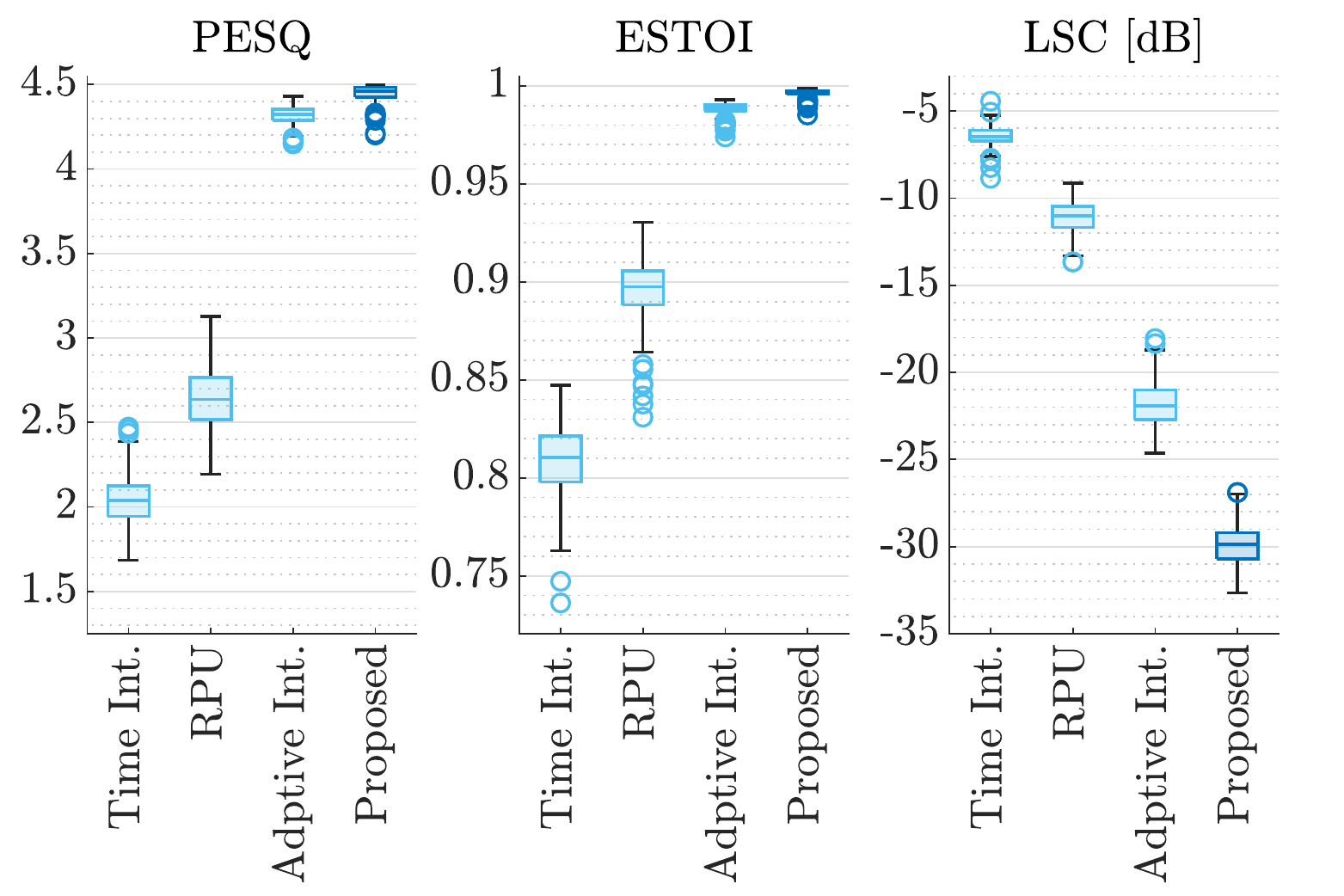}
  \caption{Boxplots of PESQ, ESTOI, and LSC of signals reconstructed from the phase differences estimated by the CNNs.
  }
  \label{fig:exp-2nd-stage-cnn}
\end{figure}

Fig.~\ref{fig:exp-2nd-stage-cnn} shows the results using the phase differences estimated by the CNNs.
According to Section~\ref{sec:exp-1st-stage}, the estimation accuracy of FPD was improved by using the CNN from that of \eqref{eq:fdiff-ave-ana}.
The performance was improved in all methods except for the time integration that does not use FPD.
The proposed optimization-based method outperformed the other methods including the adaptive integration.
This should be because the proposed method jointly optimizes all the phase at each time-frame based on the carefully designed weight.

\subsection{Comparison with DNN-based Direct Phase Reconstruction}
\label{sec:exp-vs-direct}

\begin{figure}[t!]
  \centering
  \includegraphics[width=0.99\columnwidth]{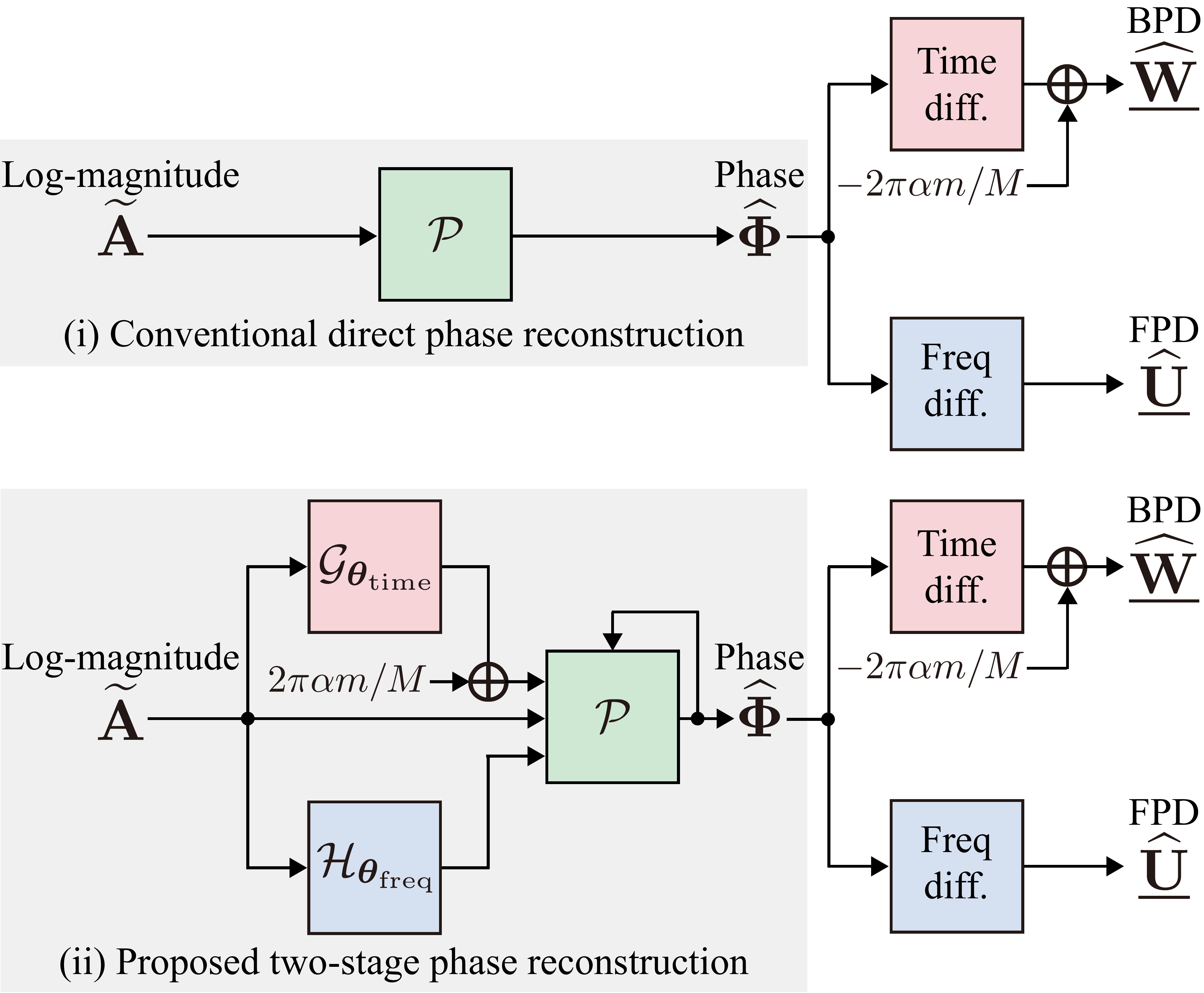}
  \caption{Block diagram of the comparison between (\romaone) conventional direct phase reconstruction and (\romatwo) proposed two-stage phase reconstruction.
  The reconstructed phase $\widehat{\boldsymbol{\Phi}}$ was evaluated through the BPD $\underline{\widehat{\mathbf{W}}}$ and FPD $\underline{\widehat{\mathbf{U}}}$ calculated from it.
  Although magnitude and phase for all time-frames are shown here, phase reconstruction was conducted in a frame-by-frame manner.}
  \label{fig:flowchart}
\end{figure}

In this experiment, the proposed two-stage framework is compared with the direct phase reconstruction using the FCN or CNN.
For the direct phase reconstruction, the DNN was trained to minimize the negative cosine loss function of phase in \eqref{eq:vm-loss}.
We further refined the estimated phase by an iterative offline phase reconstruction method called Griffin--Lim algorithm (GLA) as in the original paper~\cite{Takamichi2018}.
The number of iterations of GLA was $100$.
The reconstructed phase was evaluated by AWE of the phase differences%
\footnote{
We did not evaluate the estimated phase directly because global phase shift is not reflected in the perceptual quality.
}%
.
We also investigated the total performance of the proposed two-stage framework in this way.
To be specific, we evaluated not the phase differences estimated in the first stage but those computed from the output of the second stage as follows:
\begin{align}
    \underline{\widehat{W}}[m,n] &= \widehat{\Phi}[m,n] - \widehat{\Phi}[m,n-1] - \frac{2 \pi \alpha m}{M}, \\
    \underline{\widehat{U}}[m,n] &= \widehat{\Phi}[m,n] - \widehat{\Phi}[m-1,n].
\end{align}
This evaluation is illustrated in Fig.~\ref{fig:flowchart}.

\begin{figure}[t!]
  \centering
  \includegraphics[width=0.99\columnwidth]{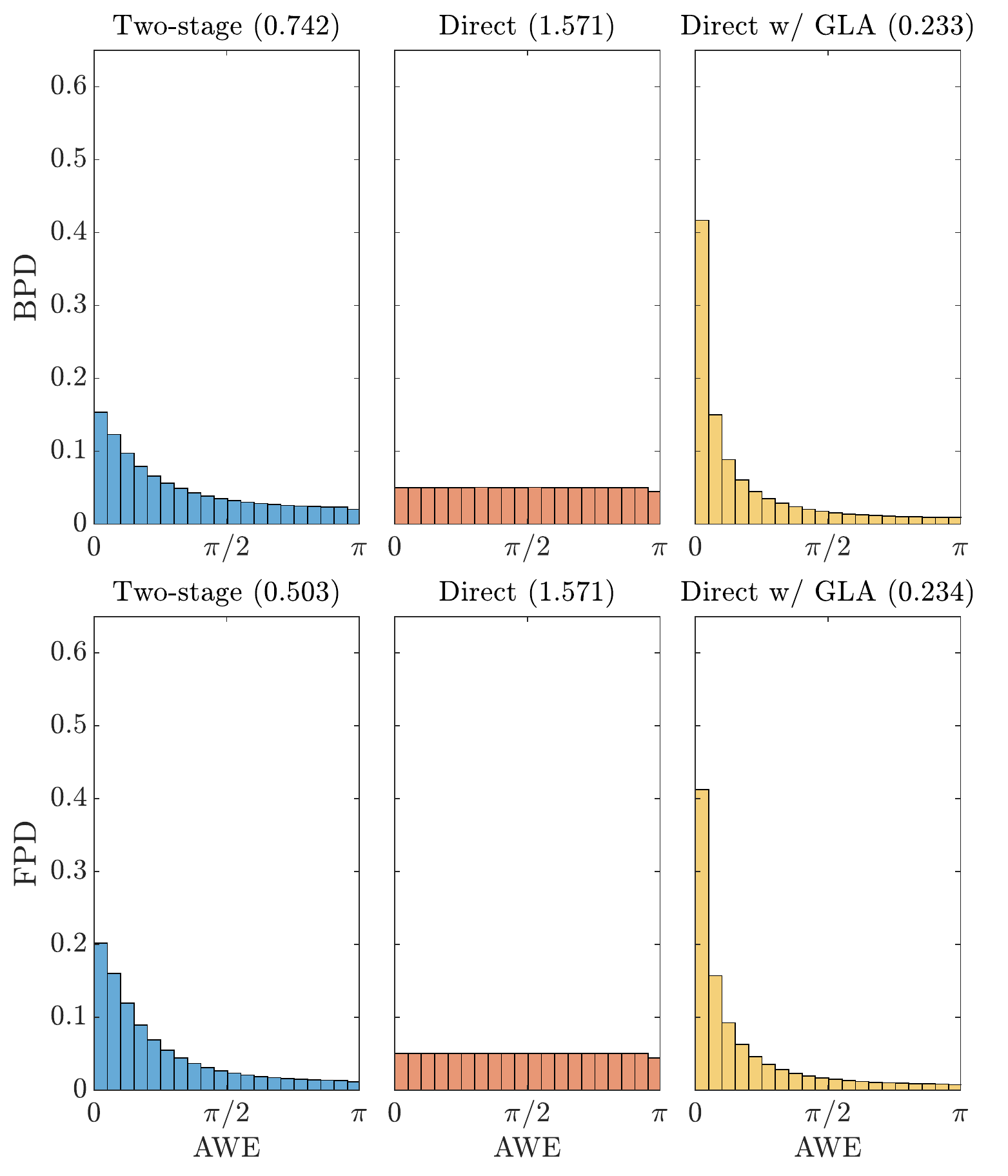}
  \caption{Histogram of AWE of the differences of reconstructed phase using FCN.
  Median over all T-F bins and audio clips is given in the parentheses.}
  \label{fig:exp-vs-direct-estimation-fcn}
\end{figure}

\begin{figure}[t!]
  \centering
  \includegraphics[width=0.99\columnwidth]{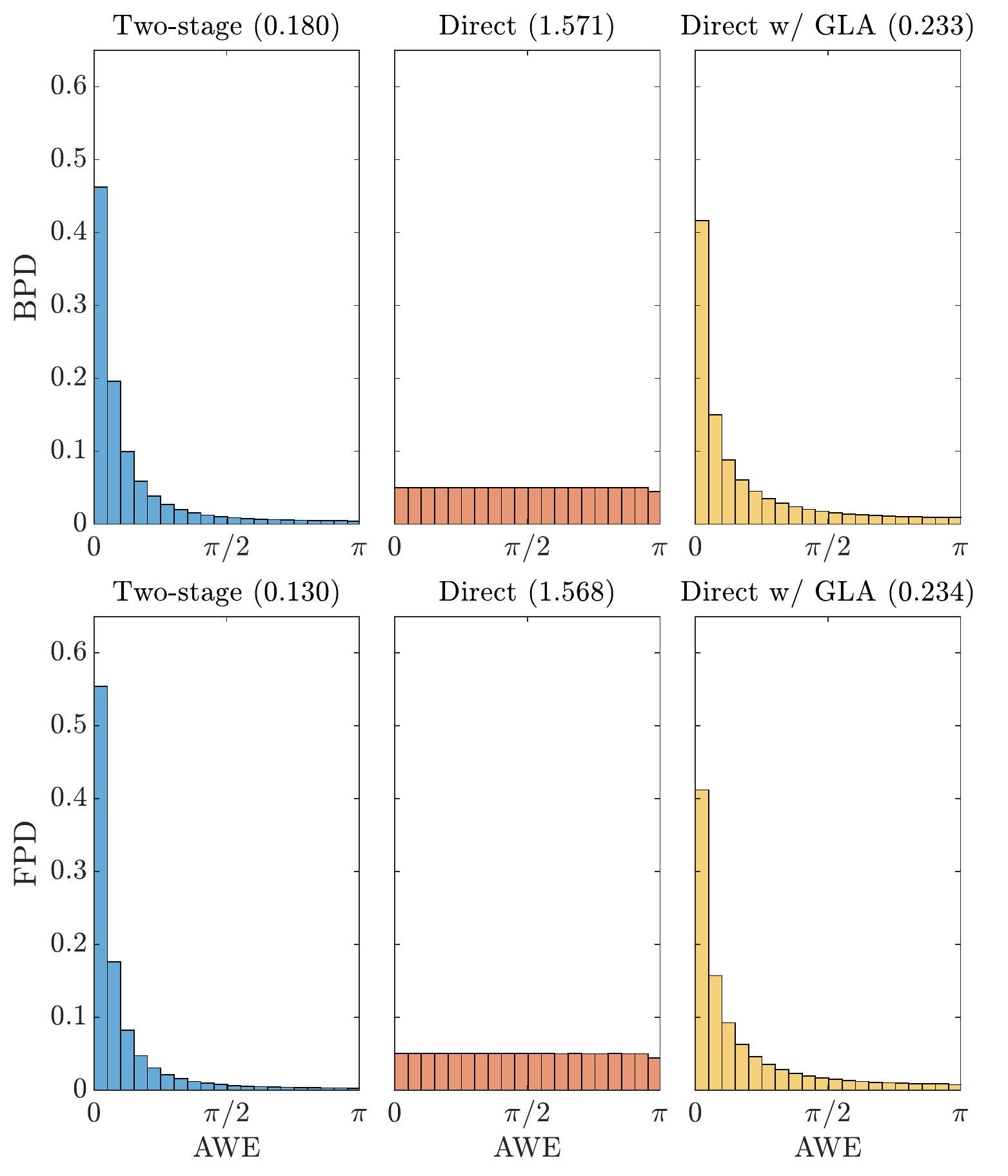}
  \caption{Histogram of AWE of the differences of reconstructed phase using CNN.
  Median over all T-F bins and audio clips is given in the parentheses.}
  \label{fig:exp-vs-direct-estimation-cnn}
\end{figure}

The results of the DNN-based phase reconstruction methods using FCNs and CNNs are summarized in Figs.~\ref{fig:exp-vs-direct-estimation-fcn} and \ref{fig:exp-vs-direct-estimation-cnn}, respectively.
The direct phase reconstruction resulted in the lowest performance regardless of the types of DNN%
\footnote{This is not due to the evaluation metric.
We confirmed that the estimated phase had large error which is reflected in BPD and FPD of the figures.}%
.
This result indicates the difficulty of directly estimating phase as discussed in Section~\ref{sec:motivation}.
In contrast, the two-stage framework with the FCNs successfully reconstructed phase up to global constant phase.
Although the direct phase reconstruction performed better when GLA was applied, the proposed framework with the CNNs outperformed it.
Note that this comparison is unfair because GLA is an iterative offline method that uses the information from all time-frames.
Even though the proposed two-stage method is non-iterative and causal, it was able to outperform the combination of the DNN-based and iterative methods.

Although the proposed two-stage method performed better than the existing methods, there is room for improvement.
According to Fig.~\ref{fig:bpd-fpd}, AWE of BPD and FPD estimated in the first stage using the FCNs were $0.417$ and $0.480$, respectively, and those using the CNNs were $0.091$ and $0.117$, respectively.
The results in Figs.~\ref{fig:exp-vs-direct-estimation-fcn} and \ref{fig:exp-vs-direct-estimation-cnn} indicates that the output of the second stage was worse than those obtained in the first stage in terms of phase differences.
This implies that the second stage of the proposed framework still has room for improvement.
Refinement of the second stage is left as a future work.

\section{Conclusion}
\label{sec:conclusion}

In this paper, we have presented a two-stage online phase reconstruction framework.
In the framework, BPD and FPD are estimated by the causal DNNs based on $1$-D frequency convolution layers.
Then, phase is reconstructed from the estimated phase differences by analytically solving the weighted least squares problem of complex STFT coefficients in a frame-by-frame manner.
We confirmed that the proposed method outperformed existing online phase reconstruction methods.
We also demonstrated the effectiveness of the two-stage framework by comparison with the direct reconstruction method.

In future works, we will apply the two-stage online framework to low-latency speech enhancement and separation.
In these applications, the noisy phase of an observed signal is useful to improve the estimation accuracy of the phase differences.
Fine-tuning of the DNNs used in the first stage to maximize the quality of the signals reconstructed by the second stage is also a possible direction for improving the proposed framework.
Together with improvement of the second stage as discussed in Section~\ref{sec:exp-vs-direct}, optimizing the whole process of the proposed framework is the next step for realizing a better phase reconstruction method.

\ifCLASSOPTIONcaptionsoff
  \newpage
\fi
\bibliographystyle{IEEEtran}
\bibliography{references_revise}

\begin{IEEEbiography}[{\includegraphics[width=1 in, height=1.25in, clip, keepaspectratio]{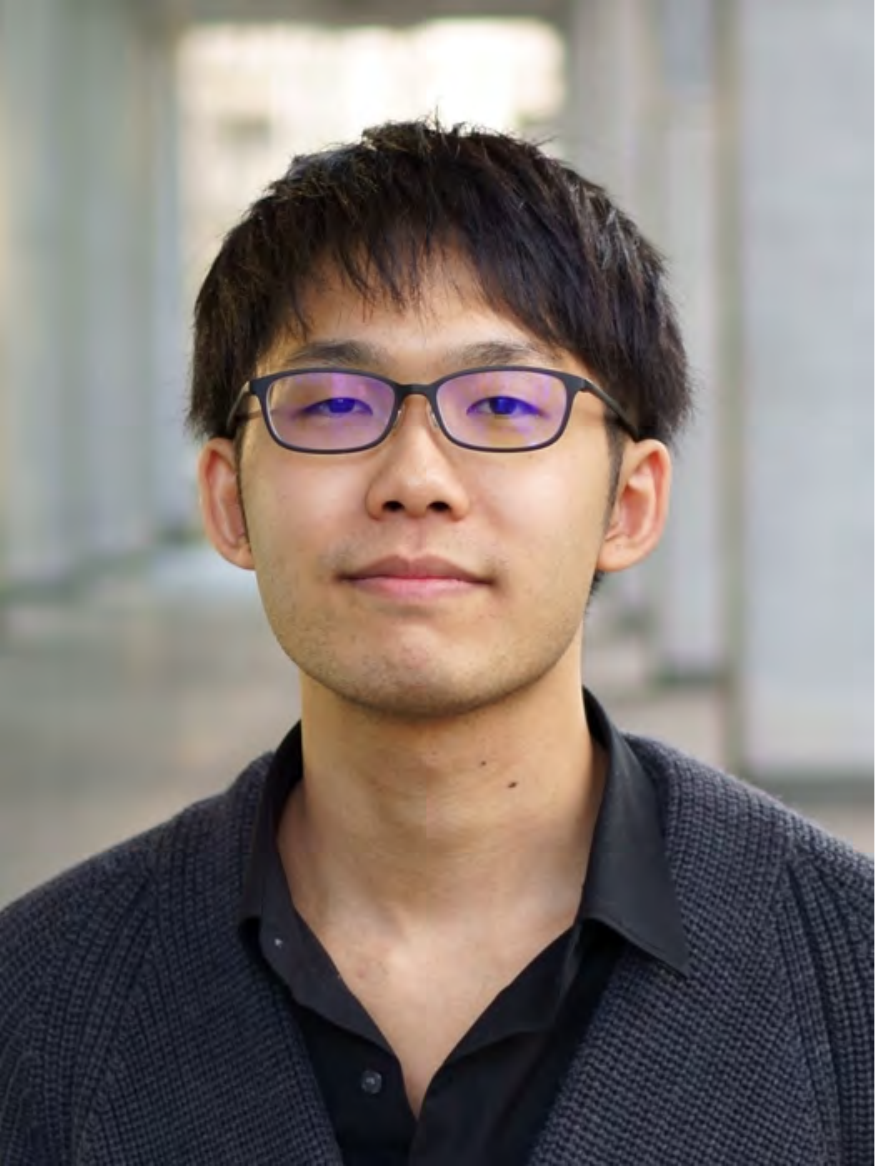}}]{Yoshiki Masuyama}
received his B.E. and M.E. degrees from Waseda University in 2019
and 2021, respectively.
He is currently pursuing the Ph.D. degree with the Graduate School of
Systems Design, Tokyo Metropolitan University.
\end{IEEEbiography}

\begin{IEEEbiography}[{\includegraphics[width=1 in, height=1.25in, clip, keepaspectratio]{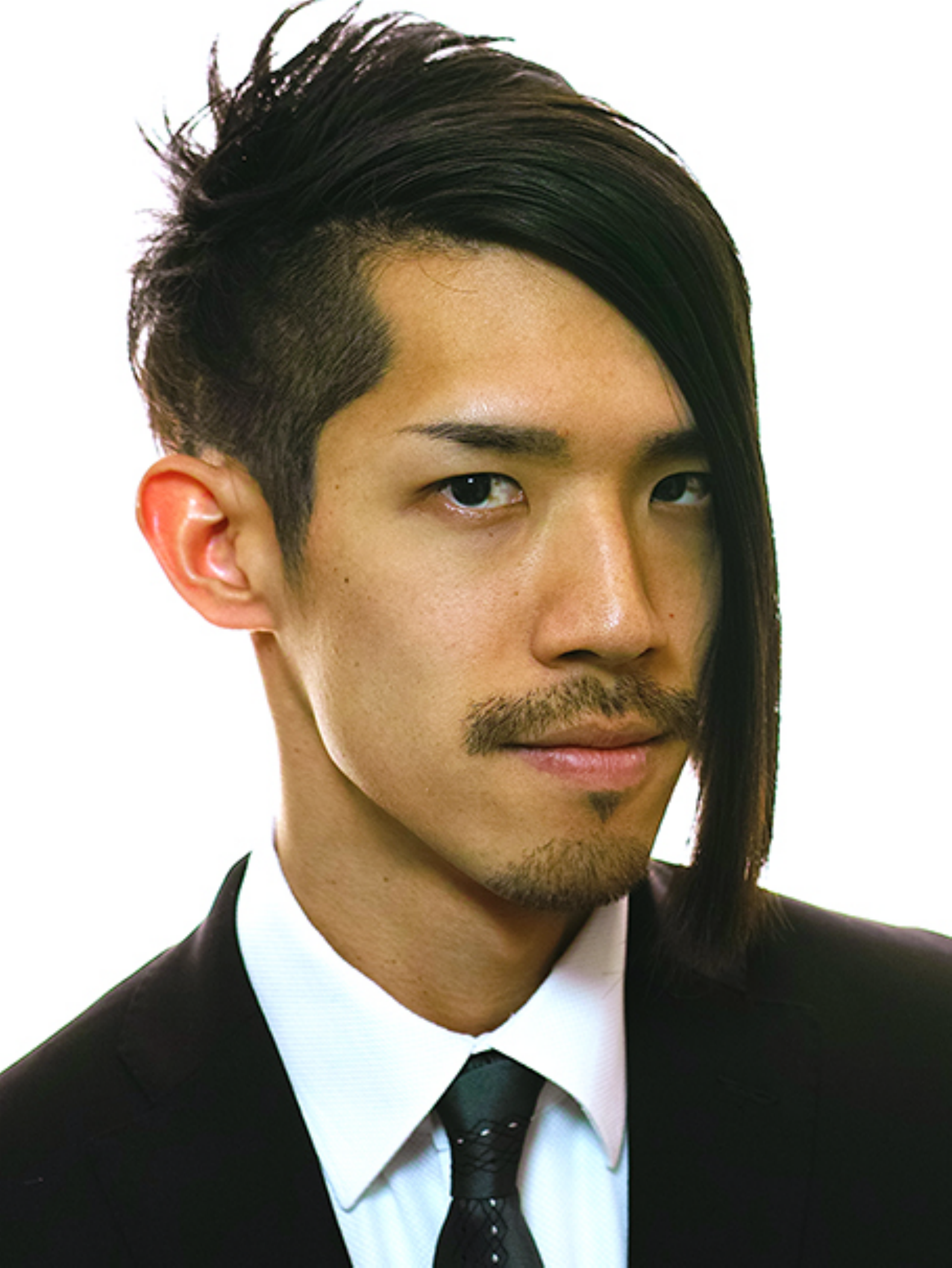}}]{Kohei Yatabe}
received the B.E., M.E., and Ph.D. degrees from Waseda University, in 2012, 2014, and 2017, respectively.
He is currently an Associate Professor with the Department of Electrical Engineering and Computer Science, Tokyo University of Agriculture and Technology.
\end{IEEEbiography}
\vfill

\begin{IEEEbiography}[{\includegraphics[width=1 in, height=1.25in, clip, keepaspectratio]{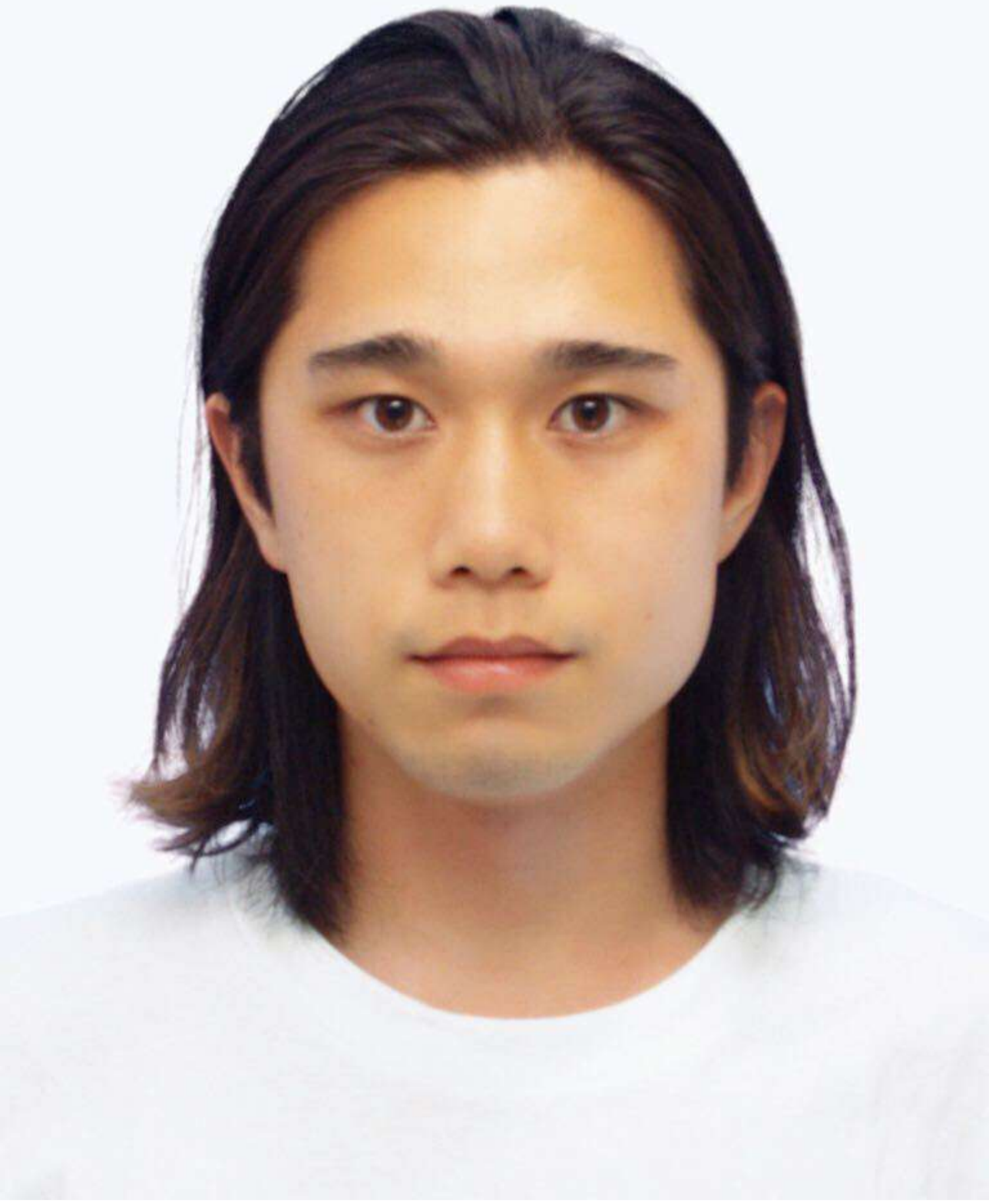}}]{Kento Nagatomo}
received the B.E. and M.S. degrees from the Department of Intermedia Art and Science,
Waseda University in 2019 and 2021.
\end{IEEEbiography}

\begin{IEEEbiography}[{\includegraphics[width=1 in, height=1.25in, clip, keepaspectratio]{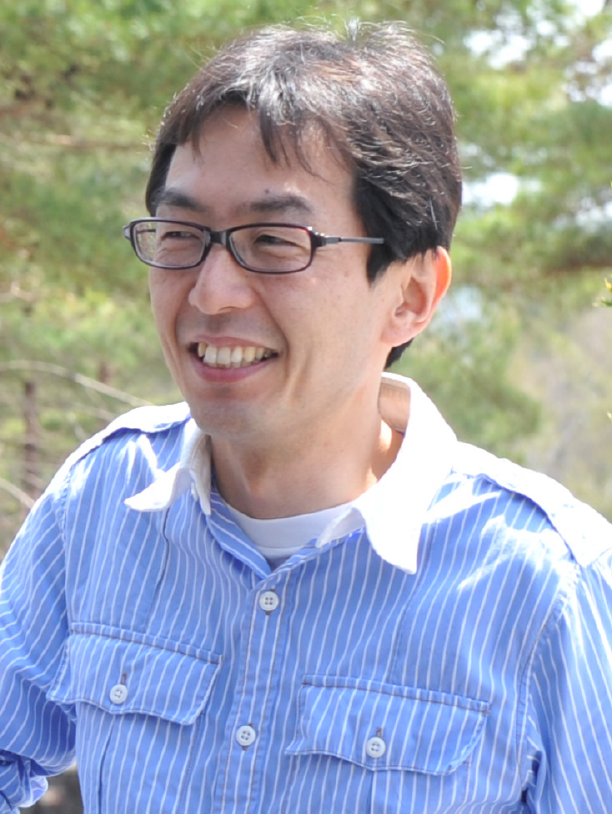}}]{Yasuhiro Oikawa}
received the B.E, M.E., and Ph.D. degrees in electrical engineering from Waseda University in 1995, 1997, and 2001, respectively.
He is currently a Professor with the Department of Intermedia Art and Science, Waseda University.
His research interests include communication acoustics and digital signal processing of acoustic signals.
He is a member of ASJ, ASA, IEICE, IPSJ, VRSJ, and AIJ.
\end{IEEEbiography}
\vfill
\end{document}